\documentclass{article}

\usepackage{arxiv}

\usepackage[utf8]{inputenc} 
\usepackage[T1]{fontenc}    
\usepackage{xr-hyper}        
\usepackage{hyperref}       
\usepackage{url}            
\usepackage{booktabs}       
\usepackage{amsfonts}       
\usepackage{nicefrac}       
\usepackage{microtype}      
\usepackage{cleveref}       
\usepackage{graphicx}
\usepackage{caption}
\usepackage{subcaption}
\usepackage{graphicx}
\usepackage{doi}

\DeclareUnicodeCharacter{03B3}{$\gamma$}
\DeclareUnicodeCharacter{0301}{\'{e}}

\title{Enhanced Patterned Fluorescence from Polystyrene through Focused Electron Beam Irradiation under Various Gases}

\usepackage{authblk}

\newbox{\orcid}\sbox{\orcid}{\includegraphics[scale=0.06]{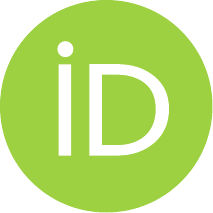}} 
\author[1]{%
	\href{https://orcid.org/0000-0001-9801-8820}{\usebox{\orcid}\hspace{1mm}Deepak Kumar\thanks{\texttt{deepak.kumar@uky.edu}}}%
}
\author[2]{%
	\href{https://orcid.org/0000-0003-2358-8006}{\usebox{\orcid}\hspace{1mm}Joseph W. Brill}%
}
\affil[1]{Department of Electrical and Computer Engineering, University of Kentucky, Lexington, Kentucky 40506}
\affil[2]{Department of Physics and Astronomy, University of Kentucky, Lexington, Kentucky 40506}
\author[1,2]{%
	\href{https://orcid.org/0000-0002-4619-7123}{\usebox{\orcid}\hspace{1mm}J. Todd Hastings\thanks{\texttt{todd.hastings@uky.edu}}}%
}

\renewcommand{\shorttitle}{Enhanced Patterned Fluorescence from Polystyrene through Focused Electron Beam Irradiation}

\hypersetup{
            pdftitle={Enhanced Patterned Fluorescence from Polystyrene through Focused Electron Beam Irradiation under Various Gases},
            pdfauthor={Deepak Kumar},
            colorlinks=true,
            linkcolor=blue,
            filecolor=magenta,      
            urlcolor=cyan,
            citecolor=cyan,
}

\begin{document}
\maketitle

\begin{abstract}
	We report on a novel method for tuning and enhancing fluorescence from irradiated polystyrene (PS) through electron-beam (e-beam) exposure in gaseous environments. We describe the effect of electron dose and ambient gas on the photoluminescence (PL) spectra and yield of irradiated PS films on insulating and conductive substrates. PS films were exposed in an environmental scanning electron microscope using a 20 keV electron beam, ambient gas pressures from $<$ 10$^{-5}$ mbar (high vacuum) to 3 mbar, and  electron doses from 1.8 to 45 mC cm$^{-2}$. Irradiated PS films were characterized using confocal microscopy, transmission electron microscopy (TEM), energy dispersive X-ray spectroscopy (EDS) and Fourier transform infrared (FTIR) spectroscopy. From emission spectra collected using confocal microscopy we found that the emission wavelength and photon yield of the irradiated film can be tuned by both dose and gas pressure.  The emission wavelength blue-shifts with increasing pressure and red-shifts with increasing dose enabling an overall tuning range of 451 – 544 nm.   Significant enhancement in the PL intensity, up to 18 times on sapphire substrates under helium when compared to high-vacuum, are observed. Overall, the highest PL yield is observed on soda lime glass substrates under argon. Also, the photon-yield on conductive substrates is significantly smaller than that yield from insulating substrates. TEM images revealed e-beam irradiated PS is amorphous in nature and elemental mapping EDS revealed no signs of film oxidation. FTIR spectroscopy revealed that under gaseous environments the decay of aromatic and aliphatic C–H stretches is reduced compared to the high vacuum exposure; in all cases, features associated with the phenyl rings are preserved. Localized e-beam synthesis of fluorophores in PS can be controlled by both dose and by ambient gas pressure. This technique could enable new approaches to photonics where fluorophores with tunable emission properties can be locally introduced by e-beam patterning.
\end{abstract}


\section{Introduction}
Fluorescent nanostructures have attracted widespread interest because of their enhanced optical properties which make them suitable for use in displays, sensors, optical information processing, labels for imaging, and diagnostics.  Radiation-induced changes to a material's optical, electrical, chemical, and structural properties offer a simple way to modify chemical and/or physical characteristics. Highly energetic particles, such as electrons or ions, deposit large amount of energy locally to bring about the aforementioned changes. Polymers are extremely susceptible to react when exposed to electron- or ion- beam radiation. Radiation synthesis of fluorophores from polymers uses electron or ion beams to generate fluorescent molecules within the polymer structure, effectively converting a non-fluorescent polymer into a fluorescent one.

\begin{figure}[!h]
    \centering
    \includegraphics[width=0.2\columnwidth]{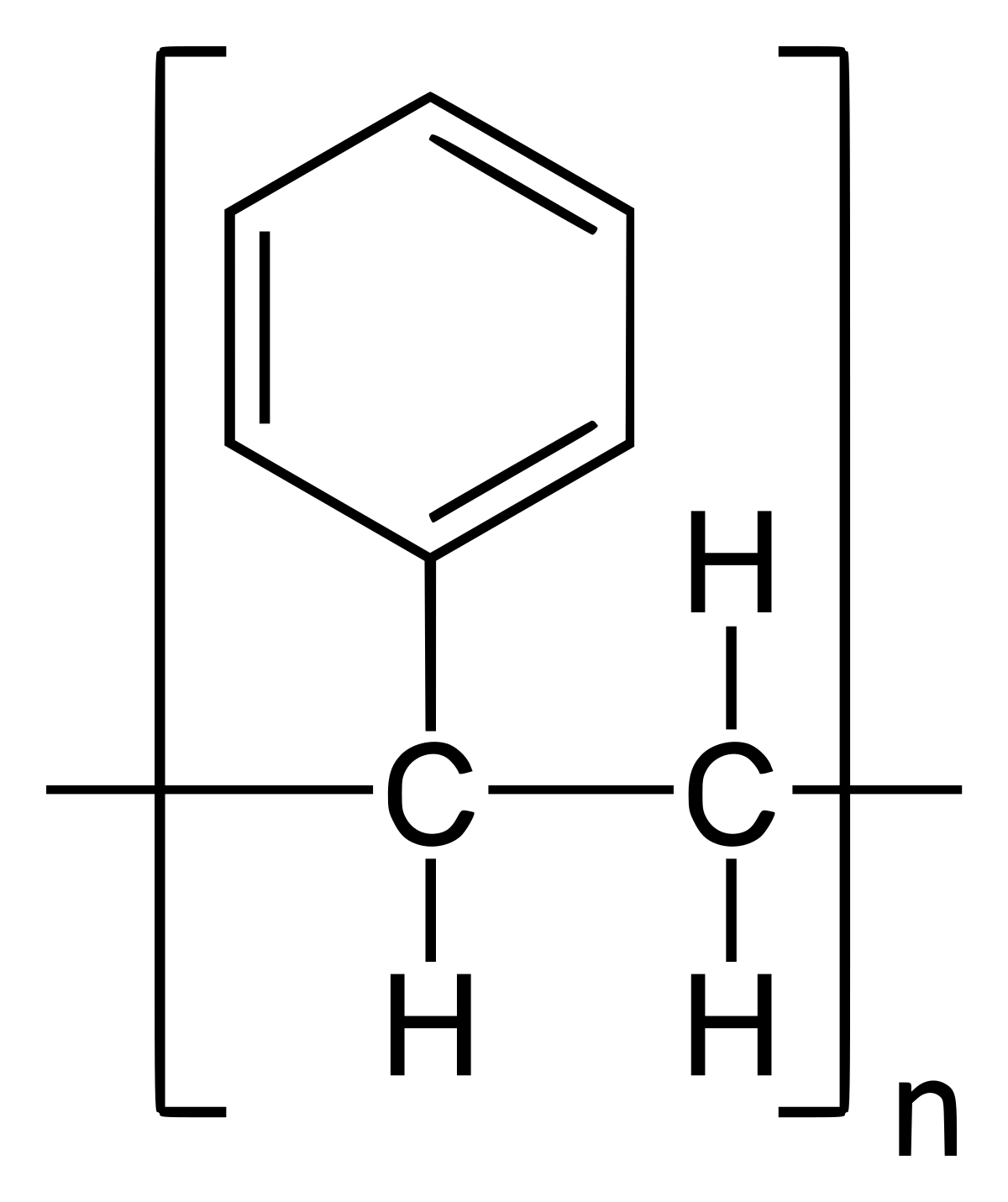}
    \caption{Chemical structure of Polystyrene}
    \label{fig:PS_Structure}
\end{figure}

Polystyrene (PS) consists of long hydrocarbon chains with a phenyl group in the repeating monomer unit (Fig. \ref{fig:PS_Structure}). PS is a negative electron-beam (e-beam) resist \cite{lai_experimental_1979, itaya_high_1982, manako_nanometer-scale_1997, manako_resolution-limit_1997, ma_polystyrene_2011, con_high_2012, con_dry_2013, dey_effect_2013} and does not exhibit luminescence in the visible spectral region \cite{gupta_time-resolved_1982}. It is known that electron irradiation transforms PS from a non-luminescent polymer into a luminescent material \cite{lee_photoluminescence_2006,lee_liquid_2007,lee_fabrication_2008,kamura_fabrication_2018,kamura_space-selective_2019}. For example, PL from irradiated PS results from formation of polycyclic aromatic hydrocarbons (PAH) or carbon dots, assumed to be composed of PAH. The effect of ionizing radiation on polymers is readily detected by the changes in molecular weight; molecular weight decreases when chain scisson is prevalent, but it increases when cross-linking is dominant. Presence of a phenyl group within the polymer molecule provides partial protection against these effects of ionizing radiation \cite{alexander_radiation_1997}. There are numerous studies investigating the range of protection offered by the phenyl radical in organic co-polymers against radiation induced cross-linking \cite{charlesby_swelling_1953, koike_radiation_1960, burlant_-radiation_1962, delides_protective_1980,pankratova_effect_2000, alexander_radiation_1997}. Significantly higher absorbed energy is required to cross-link PS, more than 2 orders of magnitude higher than many other long chain polymers \cite{charlesby_swelling_1953}, demonstrating the stabilizing effect of the phenyl group.

In the absence of a phenyl group, Randall et al. \cite{randall_13c_1983} proposed that for irradiated solid state polyethylene with absorbed dose less than the gel dose, predominant cross-linking reaction leads to the formation of Y-links. However, Horii et al. \cite{horii_carbon-13_1990} reported that H-links were predominantly present when molten state polyethylene is irradiated at higher doses. In contrast, when phenyl rings are present in the polymer chain, end links can also be formed where one or two chain scissioned reactive chain ends attach to the backbone of the second molecule \cite{noauthor_radiation_1990}. Radiation studies of p-substituted PS's indicated that the aromatic rings do not participate either in cross-linking or chain scission \cite{burlant_-radiation_1962}.   All these indicate that the reaction mechanism is significantly affected by the absorbed dose (lower or higher absorbed dose than the gel dose leads to different reaction mechanism) and also by the presence of the phenyl rings in the polymer chain. In unrelated work, water vapor has been shown to modify chemical processes during e-beam irradiation of Teflon AF \cite{jahan1993effect, forsythe1999radiation, sultan_altering_2019} and also alters the sensitivity and contrast of PMMA in e-beam lithography \cite{kumar_effect_2023}. These prior efforts motivated us to study the effect of ambient gases on the e-beam induced synthesis of fluorophores in PS.

Here we describe the effect of electron dose and gas pressure on the emission spectra and photon yield of PS films irradiated with focused electron beams on a variety of substrates. Under high vacuum exposure, we found that increasing dose red-shifts the emission spectrum and increases the photon yield, which is consistent with prior work on electron irradiated PS. However, under ambient gases, we found that the emission wavelength and photon yield can be tuned by both electron dose and gas pressure and that dramatic enhancements in photoluminescence can be achieved. This technique could enable new approaches to photonics where fluorophores with tunable emission properties can be locally introduced by e-beam patterning. 

\section{Methods}

\subsection{Spin coating}
Polystyrene (2.5K molecular weight, Scientific Polymer Products) was diluted with anisole (MicroChem Corp.) to make a 20 wt.\% solution. The PS solution thus prepared was spin coated on electrically insulating (N-BK7, soda lime glass, fused silica and sapphire) and electrically conductive (silicon and ITO coated soda lime glass) substrates at 500 rpm for 5 seconds to give a uniform layer and then spun for one minute at 2000 rpm to achieve the thickness of $\sim$600 nm. Next, the spin-coated substrate was heated on a hot plate at 70°C for 1 hour to remove any residual solvent.  A low baking temperature was chosen because low molecular weight PS thin films do not have good thermal stability \cite{ma_polystyrene_2011}. Ellipsometry (M-2000, J. A. Woollam Co. Inc.) was used to measure the film thickness of the spin-coated PS film.

\subsection{Variable pressure electron beam patterning}

\begin{figure}
    \centering
    \includegraphics[width=0.5\columnwidth]{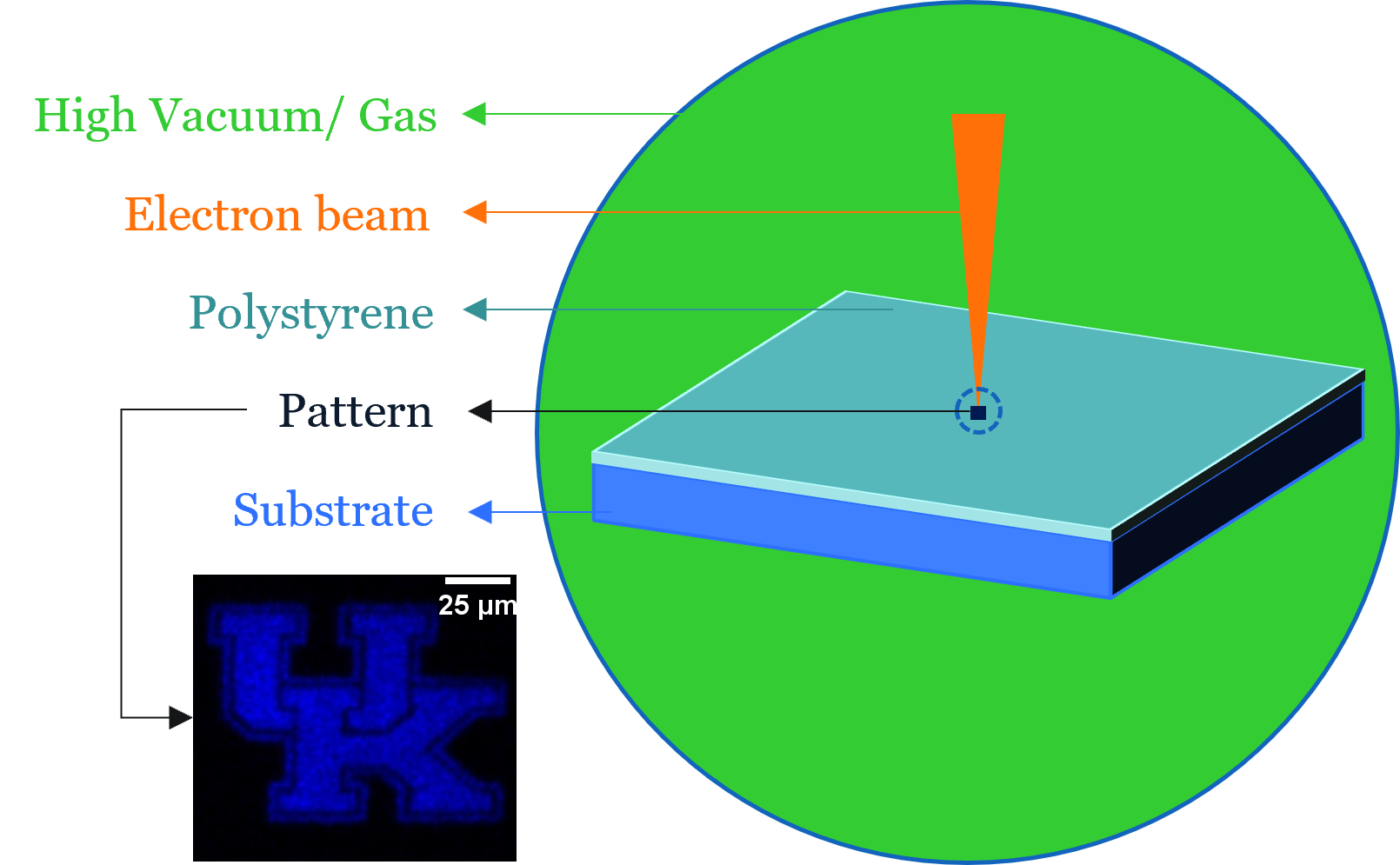}
    \caption{Schematic of electron-beam induced synthesis of fluorophores in polystyrene thin films.  The PS film is irradiated by a focused electron beam either in high vacuum or under a subatmospheric-pressure gas.  The beam is scanned to produce arbitrary fluorescent patterns, such as the one shown in the lower left.}
    \label{fig:VPEBL_Schematic}
\end{figure}

An ELPHY Plus pattern generator (Raith GmbH) coupled with a FEI environmental scanning electron microscope (Quantum FEG 250) with a fast beam blanker was used for variable pressure electron beam (e-beam) patterning process, shown schematically in Fig. \ref{fig:VPEBL_Schematic}. A working distance of 10 mm and a beam energy of 20 keV were used for all patterning processes. 100 $\mu$m square patterns were exposed with a minimum of 100 $\mu$m spacing between each neighboring square. Exposures were carried out at beam currents ranging from 0.7 to 1 nA.  Prior to each exposure process the beam current was measured under high vacuum conditions using a Faraday cup and a picoammeter (Keathley 6487).  

\subsection{Confocal microscopy}

PL measurements from irradiated patterns were carried out at room temperature using a confocal microscope (Zeiss LSM 880 Upright Multi-photon Microscope) at 405 nm laser excitation. Samples were illuminated by a focused beam with a 10x objective lens (numerical aperture 0.5) using water as immersion fluid. PL from the sample was collected by the same objective and detected by a photomultiplier (ZEISS QUASAR photomultiplier detector).  Laser power and spectral resolution were constant for all experiments.      

\subsection{Fourier-transform infrared (FTIR) spectroscopy}

Infrared reflection spectra were collected using a Thermo Fisher Scientific Nicolet 6700 FTIR spectrometer coupled to an infrared microscope. The irradiated portion of the film was completely within the illuminated region during measurements. Reflection spectra were acquired for as-coated PS and exposed patterns. 

\subsection{Transmission electron microscopy (TEM)}
PS exposed at 15 mC cm$^{-2}$ under high vacuum and 1 mbar water vapor were characterized with TEM at an accelerating voltage of 200kV (Thermo Scientific™ Talos™ F200X). Images were acquired using 16M pixel 4k x 4k CMOS camera (Thermo Scientific™ Ceta).  Elemental mapping energy dispersive X-ray spectroscopy (EDS) images were collected using Thermo Scientific™ Super-X EDS system at a beam current of approximately 0.8 nA.

\section{Results and Discussion} 

PS films show deep blue fluorescence around 410 nm upon electron irradiation \cite{lee_photoluminescence_2006}, as the electron dose varies, the PL intensity and peak wavelength also change. The PL spectrum red-shifts with increasing electron dose. At lower doses, PL intensity increases with electron dose; however, after a critical electron dose, PL intensity falls as electron dose increases.   The decrease in PL intensity beyond the critical threshold was attributed to the synthesis of new bonds in the polymer chain and the breakdown of the aromatic ring in pristine PS \cite{lee_liquid_2007}. Similar trend of decreasing PL above critical threshold is observed upon ion irradiation of polysiloxanes and polycarbosilanes derived composite ceramics \cite{pivin_photoluminescence_2000}.  

Exposing PS films to electron irradiation results in the formation of sp${^2}$ carbon structures embedded in sp${^3}$ bonds \cite{lee_photoluminescence_2006}. Irradiation causes an increase in the number of sp${^2}$ carbon clusters, the source of PL emission, below the critical electron dose. A reduction in PL intensity above the critical dose is observed as a result of either damage to the carbon clusters or defect generation in the matrix \cite{lee_liquid_2007}. In this study we conducted a methodical investigation to comprehend how the ambient gas and electron dose influence the PL spectra and yield of irradiated PS films on conductive and insulating substrates.

\begin{figure}
    \centering
    \includegraphics[width=0.75\columnwidth]{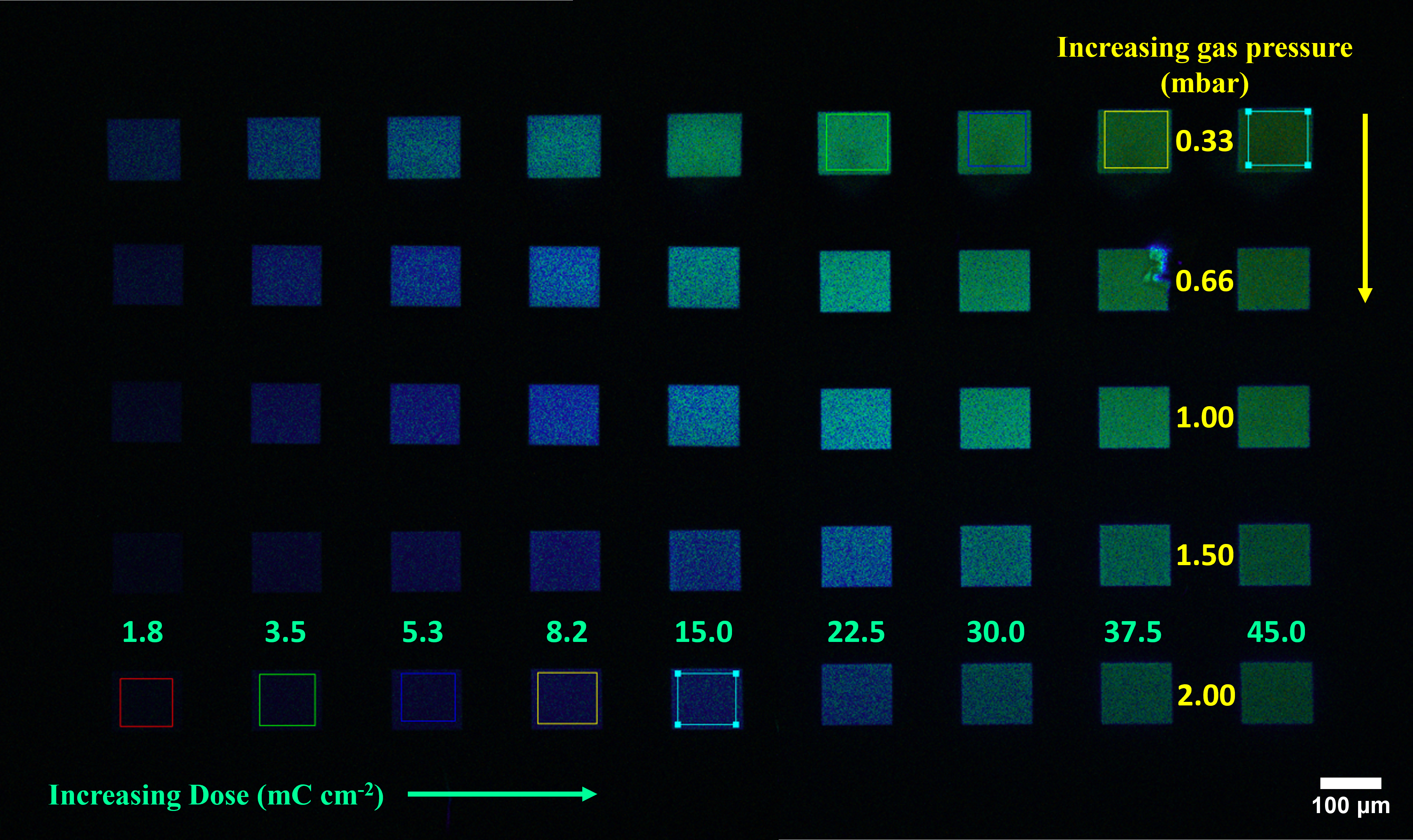}
    \caption{Fluorescence microscopy image of irradiated patterns on N-BK7 substrate. Water vapor pressure ranged from 0.33 – 2 mbar, and the dose ranged from 1.8 – 45 mC cm$^{-2}$. Exposure was done at 810 pA.  Fluorescence intensity and wavelength depend on both dose and gas pressures.}
    \label{fig:Fluorescence_microscopy}
\end{figure}

\begin{figure*}
     \centering
     \begin{subfigure}[b]{0.5\columnwidth}
         \centering
         \includegraphics[trim=65 30 92 60,clip,width=\columnwidth]{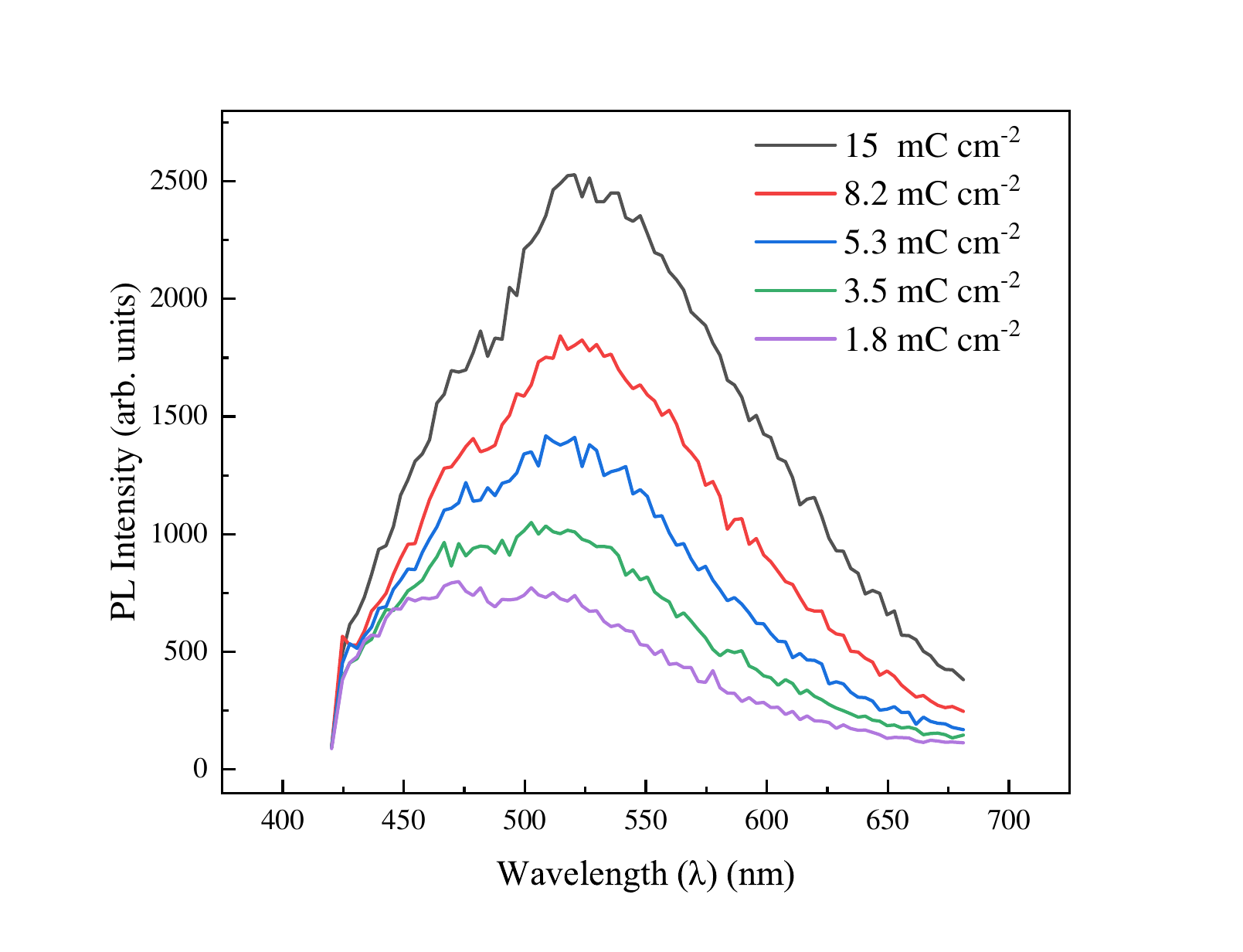}
         \caption{Exposure under high vacuum}
         \label{fig:High_Vac_on_Glass_vs_Dose}
     \end{subfigure}%
     \begin{subfigure}[b]{0.5\columnwidth}
        \centering
         \includegraphics[trim=57 30 92 60,clip,width=\columnwidth]{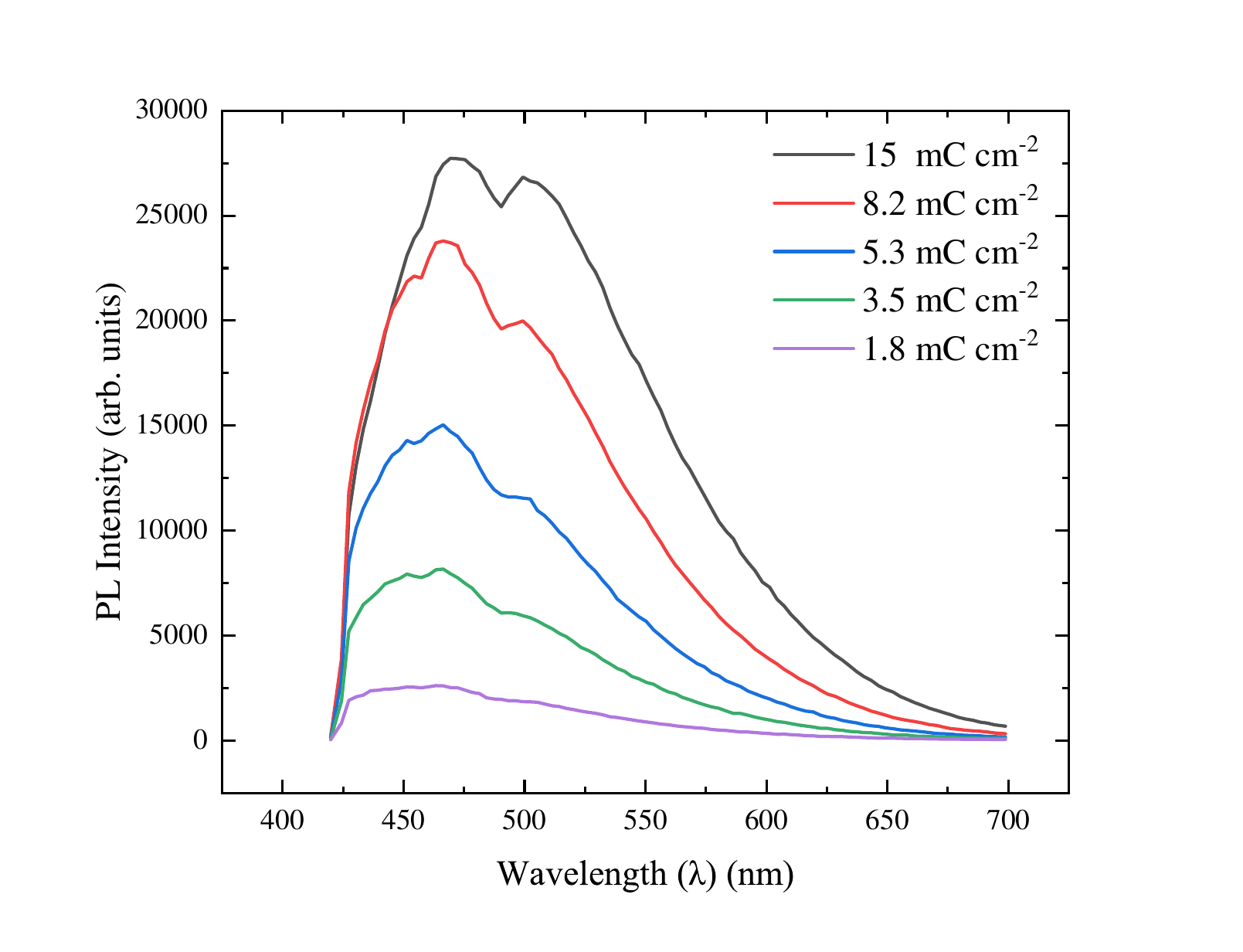}
         \caption{Exposure under 1 mbar water vapor}
         \label{fig:PL_on_glass_at_1_mbar}
     \end{subfigure}
        \caption{PL intensity from PS films on N-BK7 substrates irradiated under (a) high vacuum and (b) 1 mbar water vapor using a beam current of 692 pA.  The peak intensity is enhanced up to 10$\times$ under water vapor and the peaks are significantly sharper.}
        \label{fig:PL_on_Glass_vs_Dose}
\end{figure*}

\begin{figure}
     \centering
     \begin{subfigure}[b]{0.5\columnwidth}
         \centering
         \includegraphics[trim=57 30 92 60,clip,width=\columnwidth]{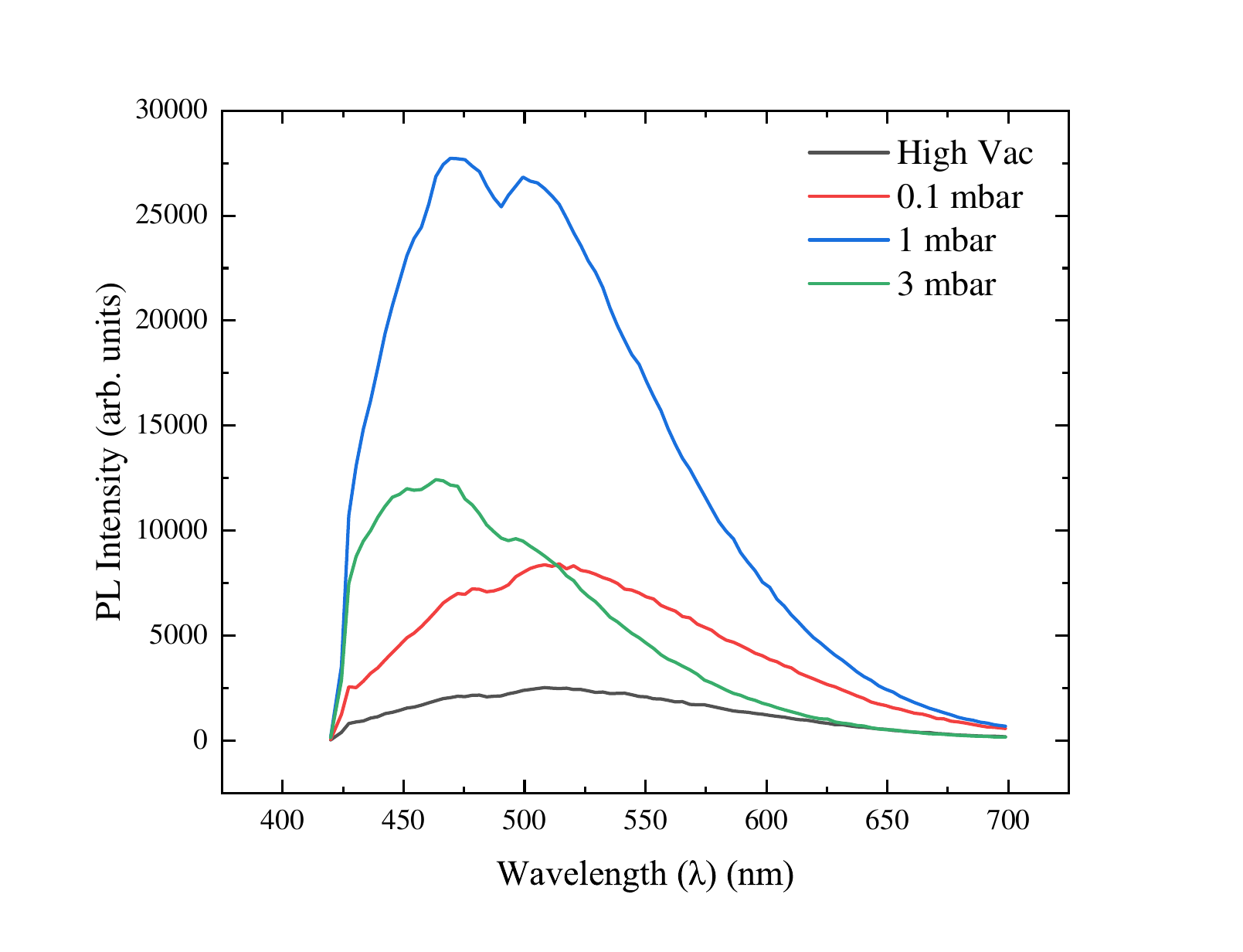}
         \caption{wide range of pressures}
         \label{fig:PL_on_Glass_wide_range}
     \end{subfigure}%
     \begin{subfigure}[b]{0.5\columnwidth}
        \centering
         \includegraphics[trim=57 30 92 50,clip,width=\columnwidth]{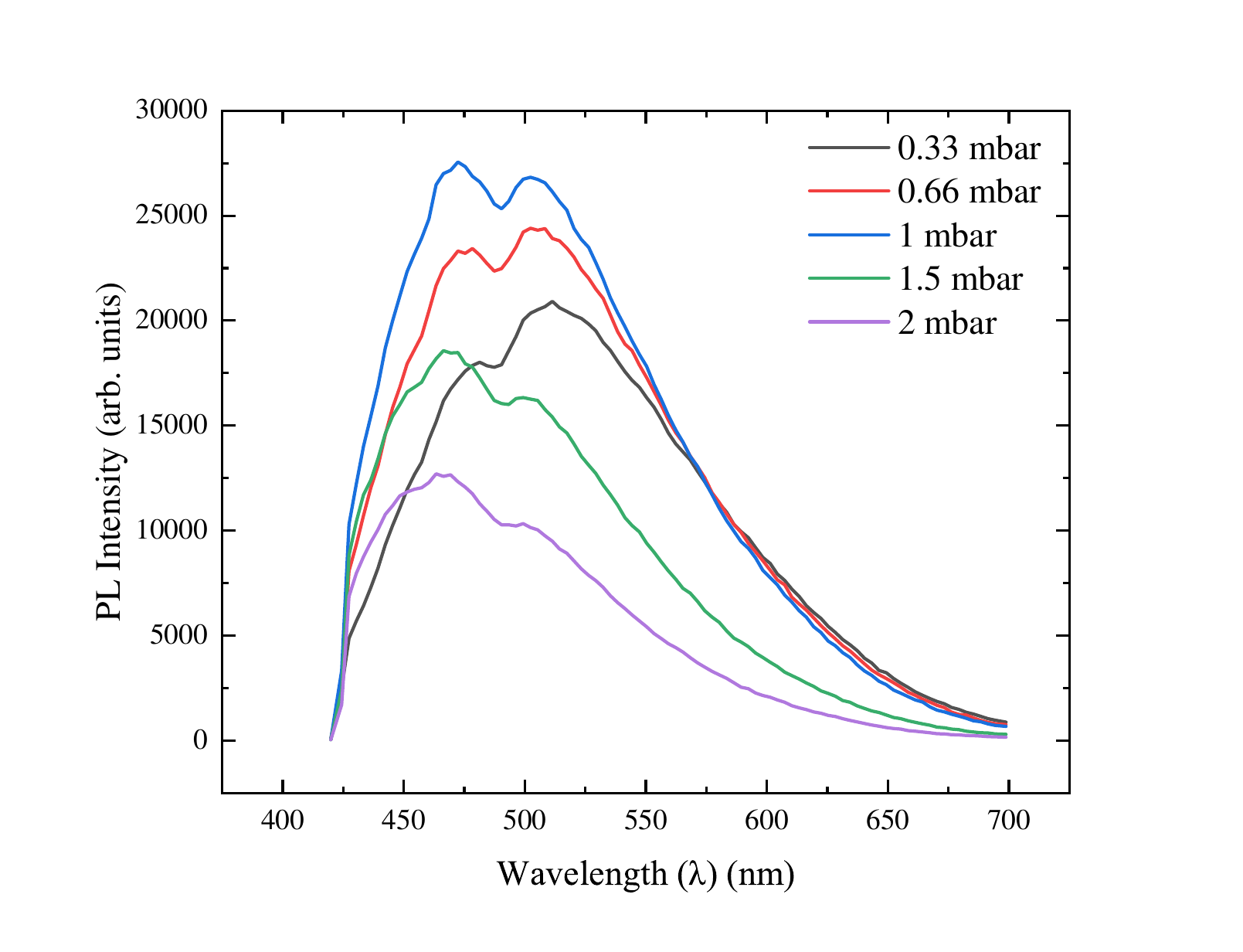}
         \caption{narrow range of pressures}
         \label{fig:PL_on_Glass_narrow_range}
     \end{subfigure}
        \caption{PL intensity as a function of water vapor pressure on N-BK7 substrate for 15 mC cm$^{-2}$ electron dose carried out at 810 pA in (a) wide range of pressures and (b) narrow range of pressures.}
        \label{fig:PL_on_Glass_vs_pressure}
\end{figure}

A fluorescence microscopy image of an example experiment is shown in Fig. \ref{fig:Fluorescence_microscopy}.  We varied the dose and water vapor pressure during e-beam patterning of PS and measured the PL from the irradiated patterns under 405-nm laser excitation. For this set of experiments, the water vapor pressure ranged from 0.33 – 2 mbar, and the dose ranged from 1.8 – 45 mC cm$^{-2}$. The pressure along rows and the doses along the columns remain the same.  We find that the emission color of the patterns can be tuned by varying the electron dose and water-vapor pressure.

\subsection{PL enhancement by e-beam irradiation of PS on N-BK7 and soda lime glass under water vapor}

Experiments were conducted to understand the dependence of electron dose and gas pressure on PL intensity. 100 $\mu$m square patterns were exposed to a focused electron beam under high vacuum and under 1 mbar water vapor in the electron dose range of 1.8 – 15 mC cm$^{-2}$. PL spectra of the patterns exposed under high vacuum, shown in Fig. \ref{fig:High_Vac_on_Glass_vs_Dose}, indicate a red-shift in the emission wavelength and an increase in the PL intensity as the electron dose increases.  These results are consistent with prior work on electron irradiated PS for the dose range considered \cite{lee_fabrication_2008, kamura_space-selective_2019}.    

Exposure under water vapor led to interesting new findings. The emission wavelength and photon yield can be tuned by both dose and water vapor pressure, as shown in Fig. \ref{fig:PL_on_glass_at_1_mbar} and Fig. \ref{fig:PL_on_Glass_vs_pressure}. The emission peak is found to blue-shift with increasing gas pressure. This could partly result from the reduction in the absorbed dose due to scattering in the gas. However, water vapor significantly \emph{increased} the photon yield, with a maximum occurring at 1 mbar pressure.  This result cannot be explained by simple electron scattering in the gas which would tend to reduce the dose with increasing pressure. The peak emission wavelength could be tuned in the 451 – 544 nm range by varying dose and water vapor pressure. The emission spectra from PS exposed under water vapor exhibit at least two peaks that are much sharper than those observed for high-vacuum exposure. This suggests that the PL in visible region originates from several PAH species in the exposed patterns.

\begin{figure}
    \centering
    \includegraphics[trim=70 30 92 60,clip,width=0.5\columnwidth]{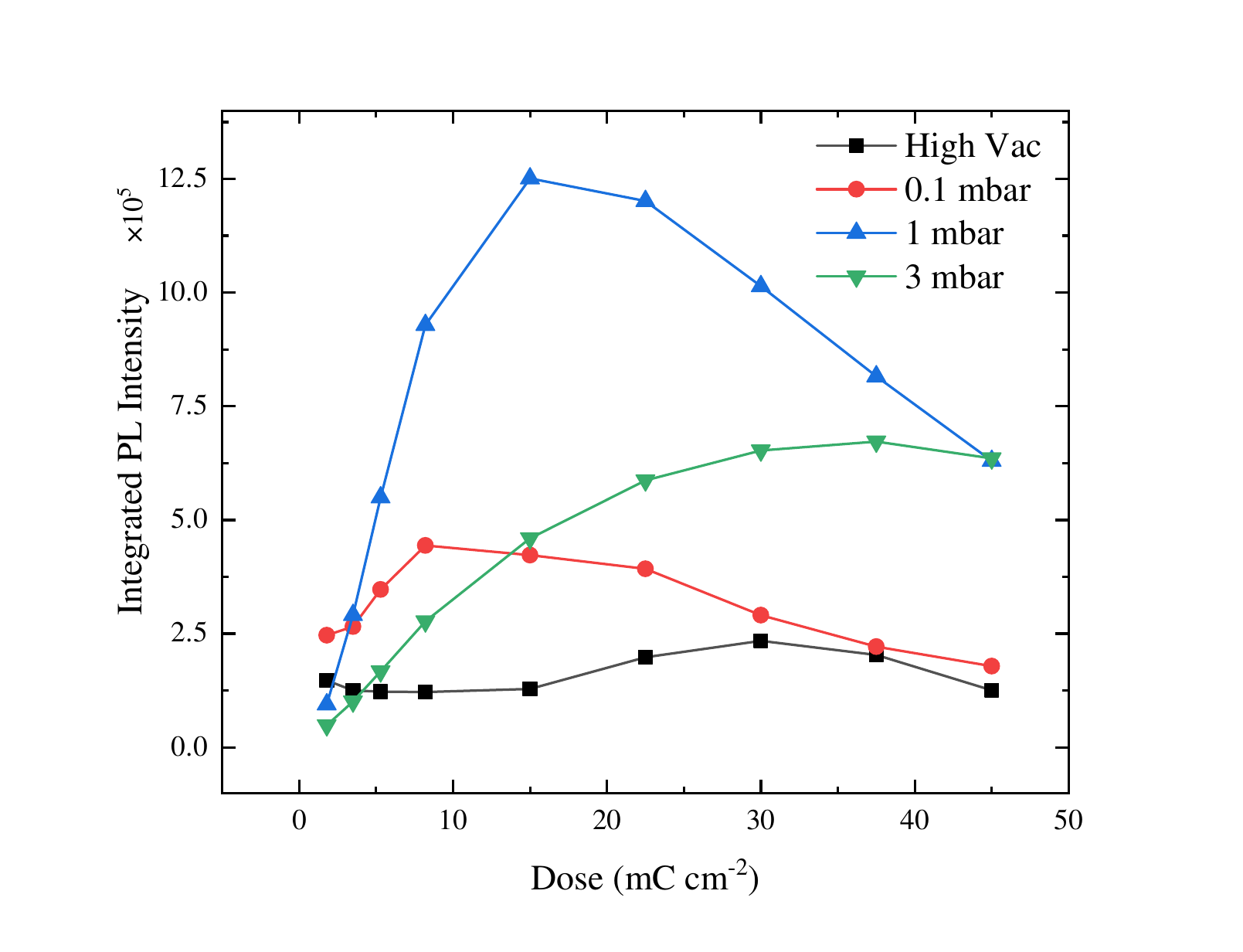}
    \caption{Integrated PL intensity as a function of electron dose for PS on N-BK7 glass irradiated under water vapor with a beam current of 810 pA.}
    \centering
    \label{fig:Intergrated_PL_Intensity_full_range}
\end{figure}

Figure \ref{fig:Intergrated_PL_Intensity_full_range} shows integrated PL intensity obtained under high vacuum and different water vapor pressures as a function of electron dose. The black line shows the integrated PL intensity under high vacuum and blue line is for 1 mbar water vapor exposure. Water vapor at 1 mbar increased the PL yield up to 10 times compared to the high vacuum exposure, a result which cannot be explained by simple electron scattering in the gas. Also, the dose at which the PL yield is maximum increases with gas pressure. The PL yield peaks around 8  mC cm$^{-2}$ for 0.1 mbar exposure, where as it peaks at 37 mC cm$^{-2}$ for 3 mbar exposure. For 1 mbar water vapor it peaks at 15 mC cm$^{-2}$ exposure dose. With increasing electron dose the PL yield peaks and gradually decreases after a critical threshold electron dose is reached. Thus, water vapor at the right pressure dramatically increases PL yield and electron scattering in gas alone cannot explain the enhanced PL. 

\subsubsection{Dependence of PL on beam current on N-BK7 substrate: }
\label{beam_current_dependence}
Fluorophore synthesis could depend on electron-beam current through at least three mechanisms.  First, if reactions involve the ambient gas, the gas could be depleted at higher beam currents similar to the effects seen in electron-beam induced deposition and etching\cite{huth2012focused}.  Second, if the gas provides incomplete charge dissipation or screening, higher beam currents could lead to increased substrate charging.  Finally, increased beam currents could increase the local temperature of the irradiated region.  
The fluorescence properties of CDs fabricated by pyrolysis \cite{krysmann2012formation}, hydrothermal \cite{song2015investigation, zhu2013highly} and ultrasonic synthesis \cite{sun2013hair} have been found to be affected by the reaction temperature, where a decrease in the PL yield with increasing reaction temperature was observed. The decrease in PL yield with increasing reaction temperature was attributed to consumption of the molecular fluorophores by the carbon core. More recently Zhang et. al \cite{zhang2016effect} demonstrated that there is an optimum reaction temperature where the PL yield reaches the threshold value. Below the optimal reaction temperature an increase in PL yield with increasing reaction temperature is observed, which was suggested to result from an increase in the number of fluorescent polymer chains. Beyond the optimal reaction temperature, the growth of the carbon core consumes the fluorescent polymer chains resulting in reduced PL yield. 

\begin{figure}
     \centering
     \begin{subfigure}[b]{0.5\columnwidth}
         \centering
         \includegraphics[trim=57 30 87 50,clip,width=\columnwidth]{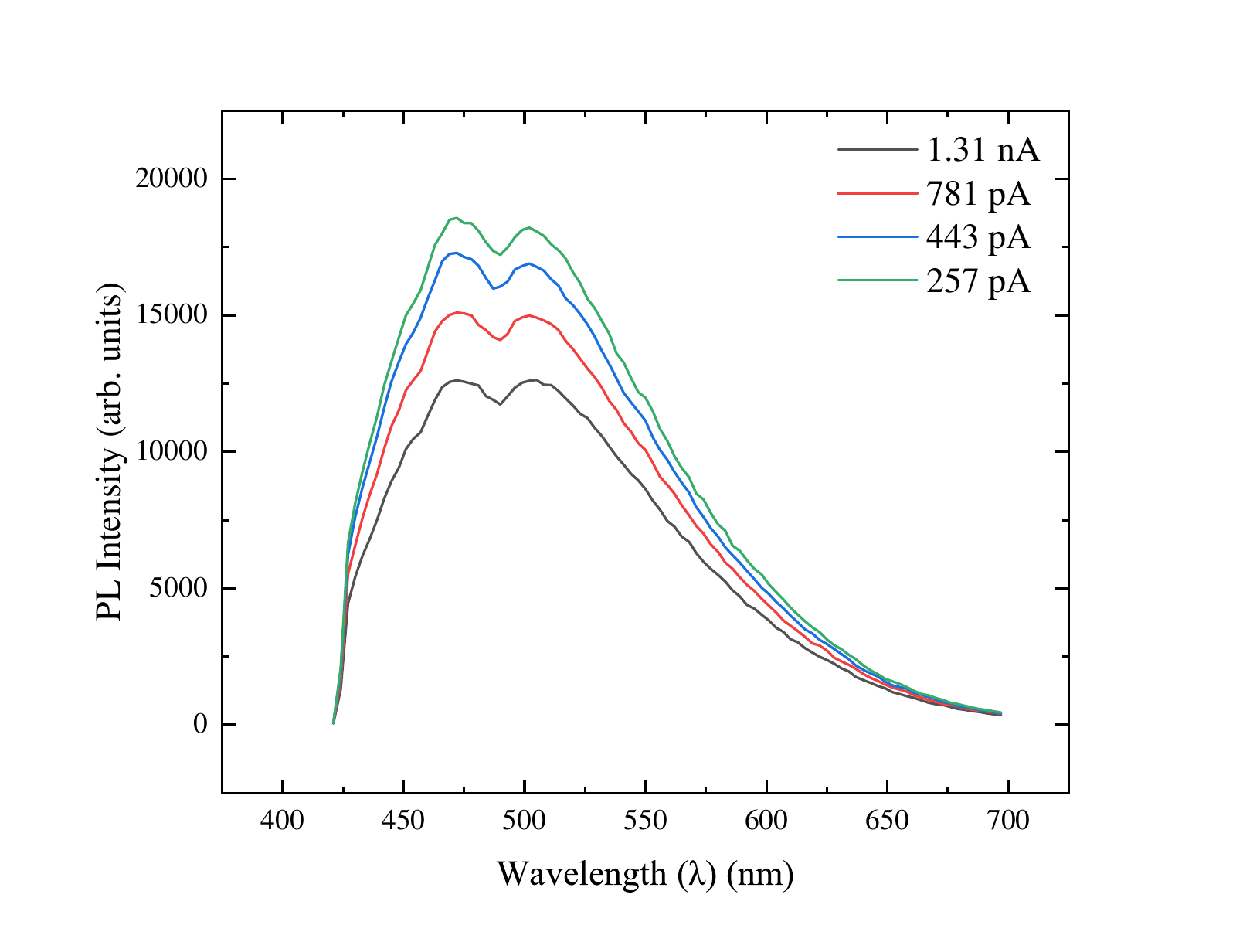}
         \caption{PL Intensity vs beam current}
         \label{fig:PL_vs_Beam_Current}
     \end{subfigure}%
     \begin{subfigure}[b]{0.5\columnwidth}
        \centering
         \includegraphics[trim=75 30 90 50,clip,width=\columnwidth]{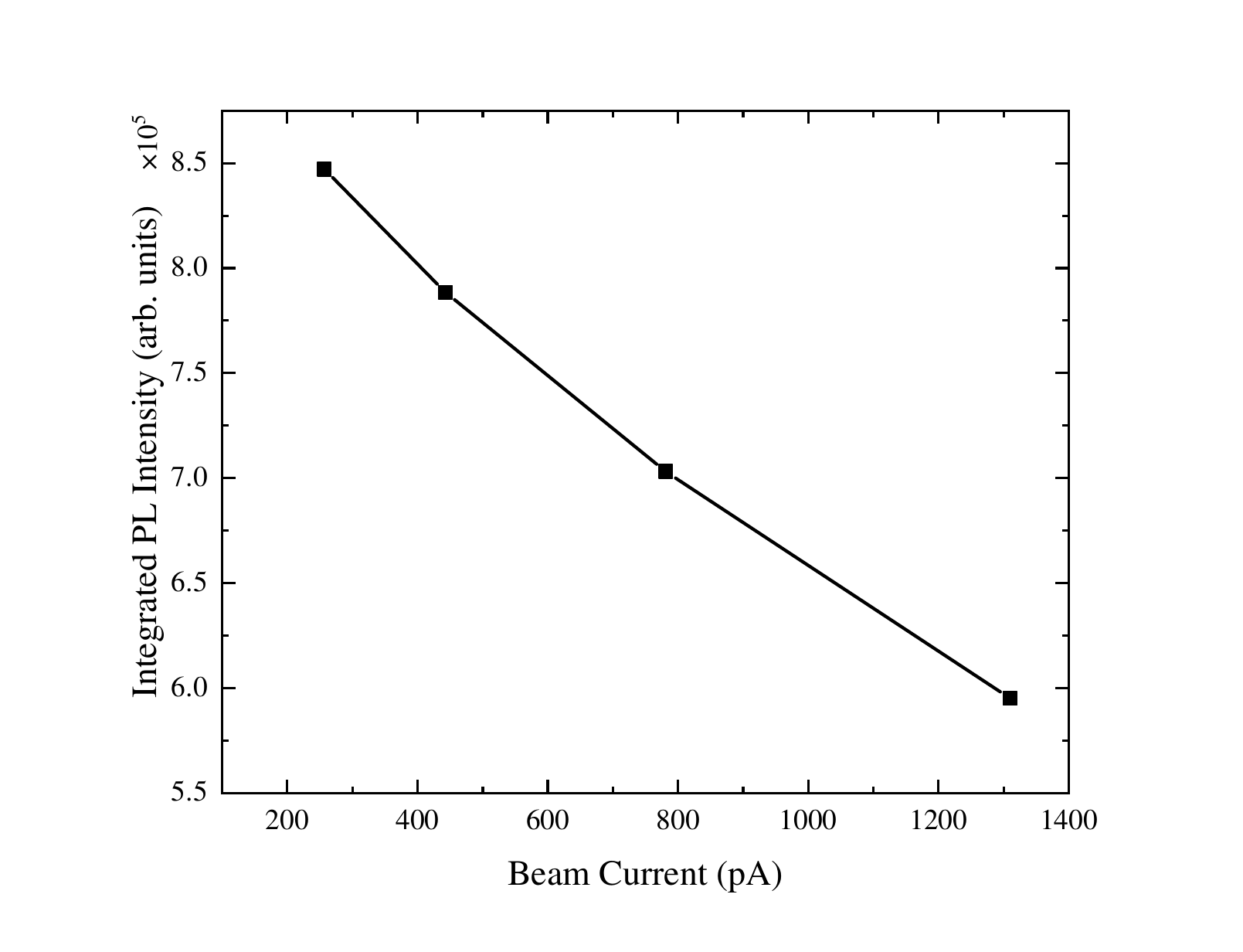}
         \caption{Integrated PL Intensity vs beam current}
         \label{fig:Int_PL_vs_Beam_Current}
     \end{subfigure}
        \caption{Beam current dependence on PL under 1 mbar water vapor on N-BK7 substrate; (a) PL intensity vs beam current and (b) Integrated PL intensity vs beam current.}
        \label{fig:Beam_Current}
\end{figure}

To investigate the effect of beam current on the PL yield of the fluorophores synthesized by e-beam irradiation, patterns were exposed with increasing beam currents under 1 mbar water vapor pressure at an electron dose of 15 mC cm$^{-2}$. It is found that the PL intensity decreases with increasing beam current (Fig. \ref{fig:PL_vs_Beam_Current}), and this decrease is approximately linear over the range of currents considered (\ref{fig:Int_PL_vs_Beam_Current}).  The reduction of PL with beam current is consistent with the effect of temperature on PL yield of CDs synthesized by other methods \cite{krysmann2012formation, sun2013hair, zhu2013highly, song2015investigation}.  However, such a result could also be consistent with sufficient substrate charging to repel electrons and reduce the local dose and with depletion of the gas available for synthesis reactions.  Thus, further investigation is required to isolate the underlying mechanism of beam-current dependence.

\subsubsection{N-BK7 vs Soda lime glass substrate under 1 mbar water vapor:}
\label{N-BK7_vs_SLG}

\begin{figure}
    \centering
    \includegraphics[trim=90 30 90 50,clip,width=0.5\columnwidth]{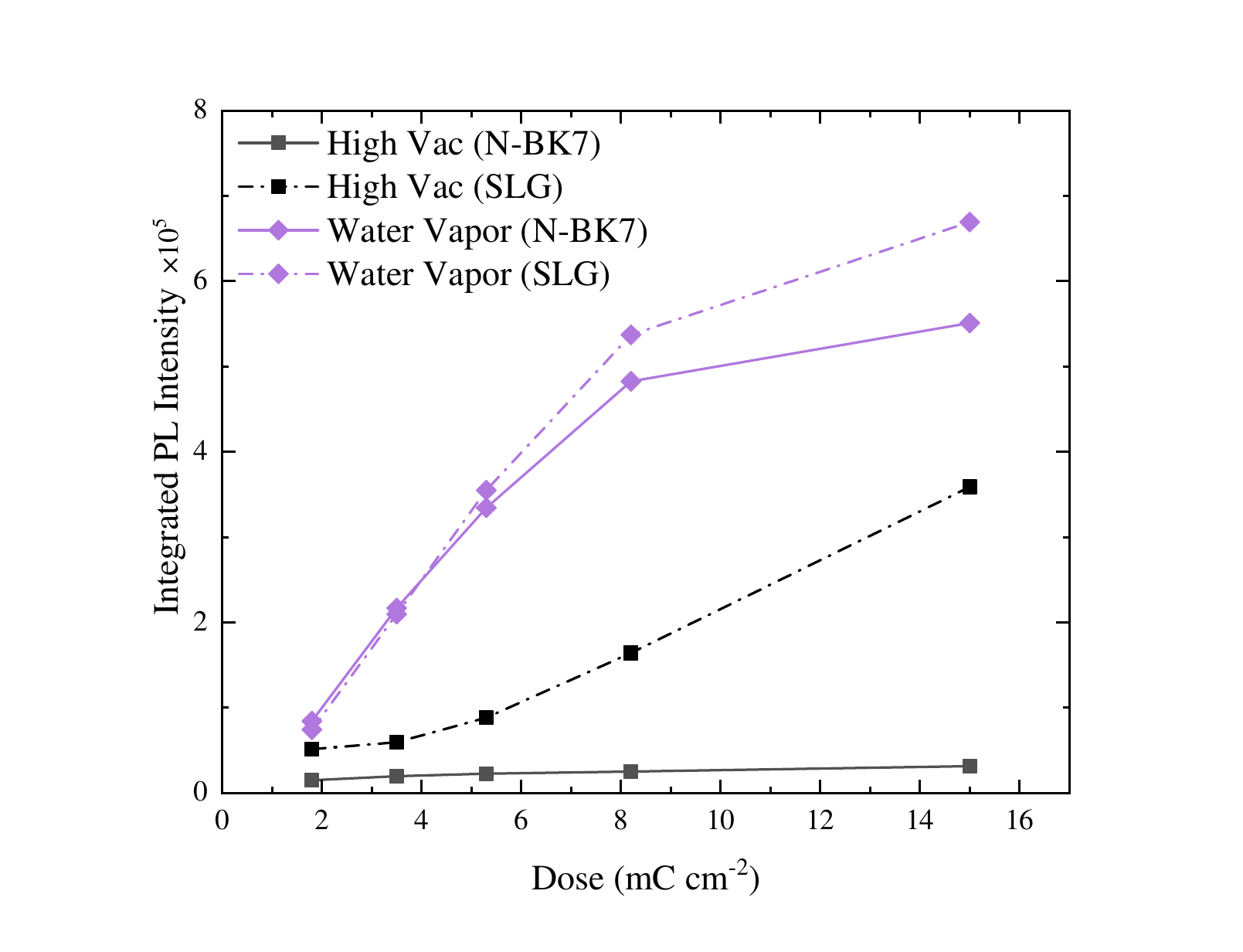}
    \caption{N-BK7 vs soda lime glass: Integrated PL intensity as a function of exposure dose under 1 mbar water vapor. N-BK7 and soda lime glass substrate were exposed at a beam current of 764 pA and 950 pA respectively.}\centering
    \label{fig:IPL_N-BK7_vs_SLG}
\end{figure}

To further understand the effect of water vapor on the PL intensity, we compared integrated PL intensity for patterns exposed under high vacuum and 1 mbar of water vapor pressure as a function of electron dose on N-BK7 and soda lime glass substrate as shown in Fig. \ref{fig:IPL_N-BK7_vs_SLG}. One of the primary differences between N-BK7 and Soda lime glass lies in their chemical composition. The presence of mobile Na$^+$ ions in soda lime glass makes it effective for charge dissipation under high electric field. Glasses containing alkali ions upon focused electron irradiation undergo significant structural changes due to the formation of high electric field created by the trapped electrons inside the exposed volume making the migration of ions easier until the ions recombines with a free or trapped electrons \cite{gedeon1999fast, gedeon2000microanalysis}, thus, providing a charge dissipation pathway. 

Another difference between the two substrates is that the thermally conductivity of N-BK7 is slightly higher than that of soda lime glass, see Table. \ref{substrate_thermal_conductivity}. The reaction temperature of the exposed region on soda lime glass substrate is also expected to be slightly higher than on the N-BK7 substrate due to the lower thermal conductivity. The integrated PL intensity from irradiated PS on soda lime glass is found to be greater than on N-BK7 for exposure under high vacuum as well as under 1 mbar water vapor.  This is consistent with  dose reduction by greater charging of N-BK7.  It could also be consistent with more efficient   indicating that the increase in PL yield with increasing reaction temperature, below the optimum temperature, resulting from an overall increase in the number of fluorescent polymer chains \cite{zhang2016effect}. 

Thus, differences in the electrical and thermal properties of the substrates could lead to different reaction mechanisms for fluorophore formation resulting in different PL characteristics for exposures under otherwise identical conditions. We present a detailed analysis of PL enhancement on substrates with different electrical and thermal properties later in section \ref{Insulating_vs_conducting_substrates}. 

\subsubsection{Transmission electron microscopy for patterns exposed on soda lime glass substrates: }

\begin{figure}
     \centering
     \begin{subfigure}[b]{0.45\columnwidth}
        \centering
        \includegraphics[width=\columnwidth]{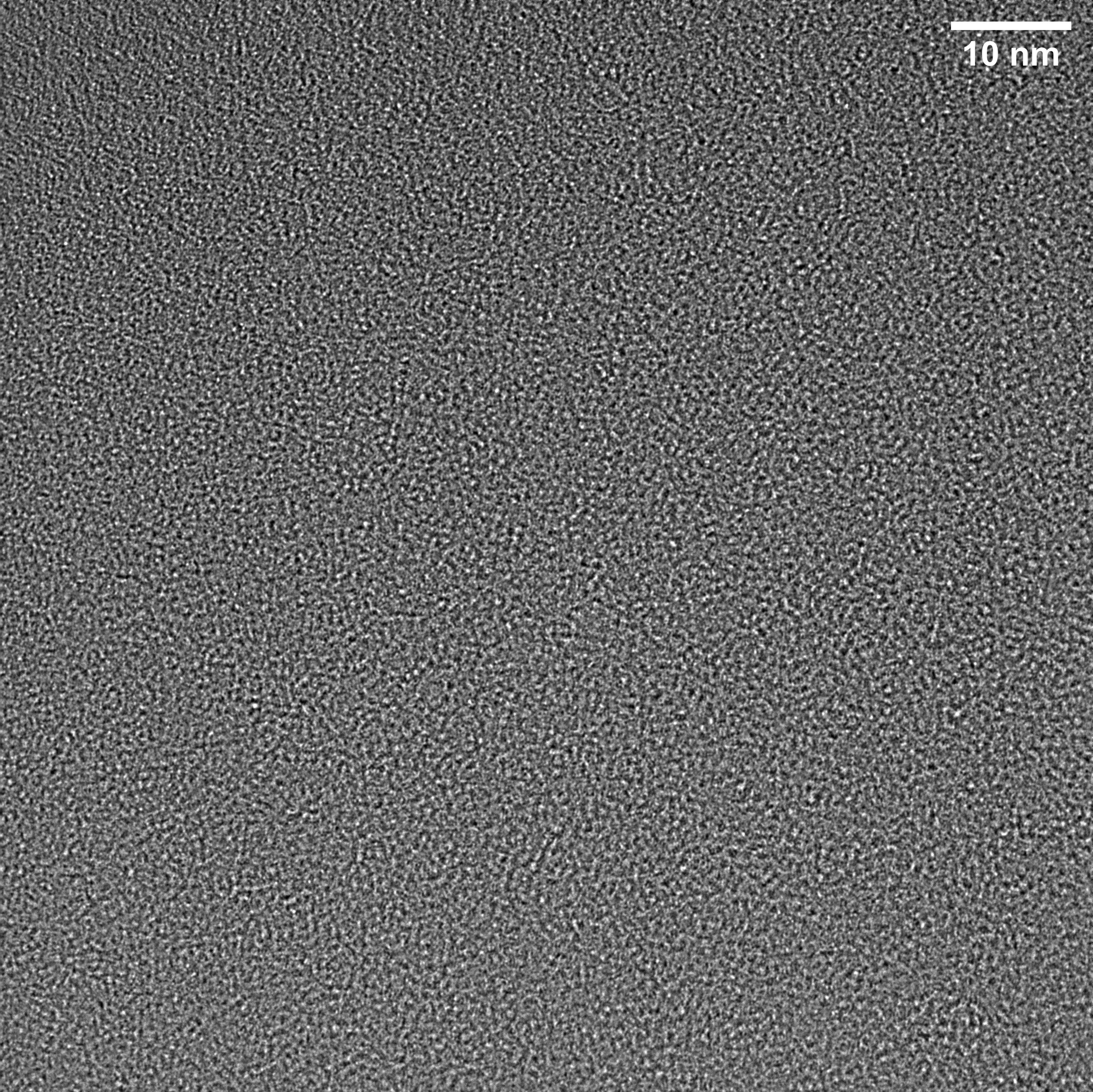}
        \caption{1 mbar water vapor}
         \label{fig:TEM_H2O}
     \end{subfigure}
     \hspace{1cm}
     \begin{subfigure}[b]{0.45\columnwidth}
        \centering
         \includegraphics[width=\columnwidth]{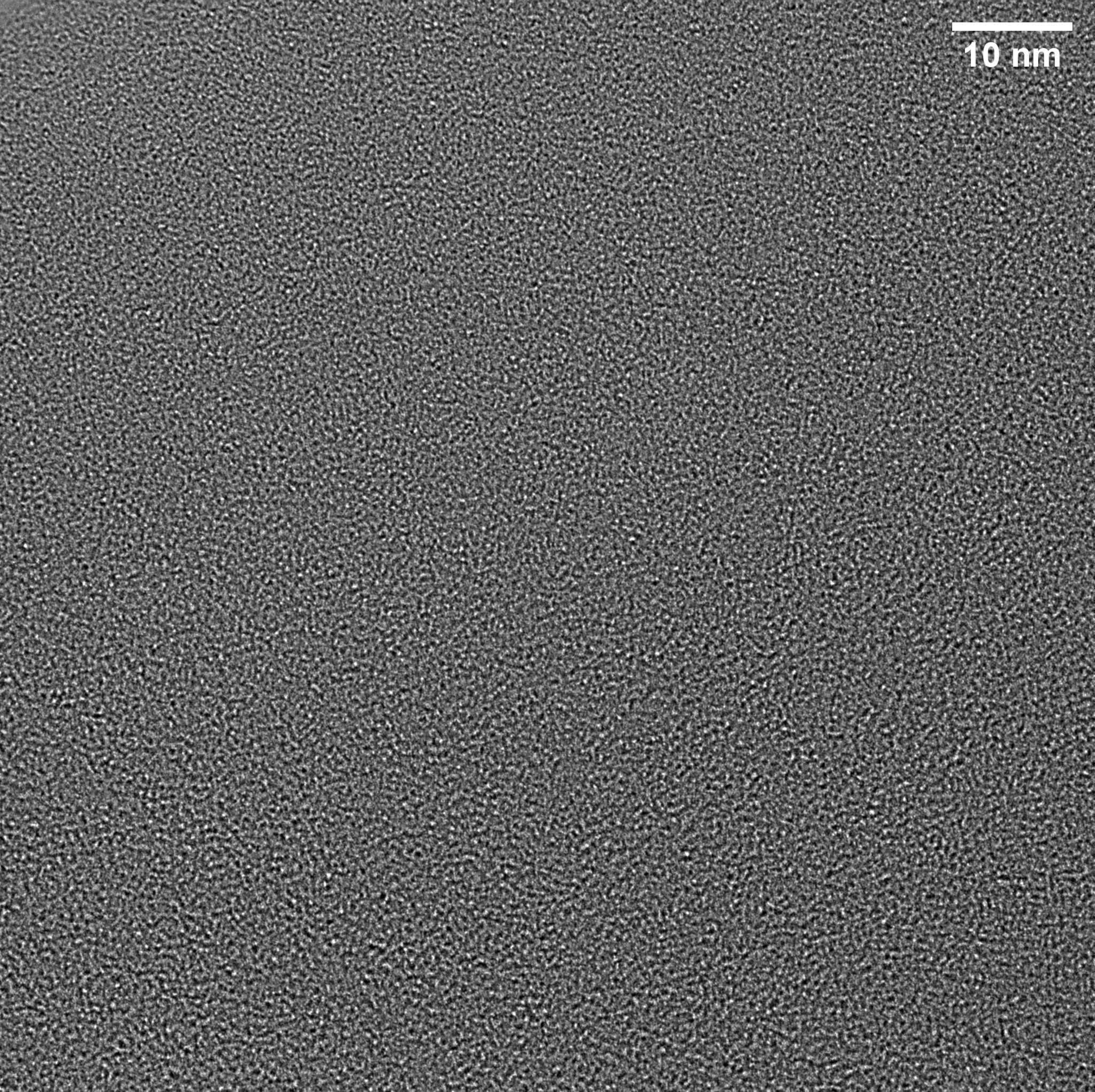}
         \caption{High vacuum}
         \label{fig:TEM_High_Vac}
     \end{subfigure}
        \caption{TEM image of  PS film on soda lime glass substrate irradiated at 15 mC cm$^{-2}$ under (a) 1 mbar water vapor and (b) high vacuum.}
        \label{fig:TEM}
\end{figure}

High-resolution TEM image of e-beam irradiated PS (on soda lime glass substrate) under 1 mbar water vapor at an electron dose of 15 mC cm$^{-2}$, shown in Fig. \ref{fig:TEM_H2O}, indicates that the irradiated film is amorphous in nature with no obvious lattices. In our experiments we found that even for the high vacuum exposure the irradiated films are amorphous in nature, Fig. \ref{fig:TEM_High_Vac}, as opposed to the crystalline nature of CDs obtained by similar e-beam exposure of PS \cite{kamura_space-selective_2019}.   

\subsubsection{Energy dispersive X-ray Spectroscopy mapping for patterns exposed on soda lime glass substrates: }

The nature of CDs' optical characteristics has been the subject of numerous investigations. While some studies have claimed that certain CD subsets gives rise to the optical characteristics \cite{ghosh_photoluminescence_2014}, others have linked these to states originating from various functional groups \cite{dong_carbon-based_2013}. One frequently made hypothesis is that the primary absorption properties are derived from core states \cite{nie_carbon_2014, wen_intrinsic_2013, yu_temperature-dependent_2012, das_single-particle_2014}. The luminescent property of PS nanospheres formed by pulsed ultra violet laser radiation in air and vacuum, revealed the production of carbonyl groups on the surface of PS by photochemical oxidation is what causes the white luminescence \cite{kim_white_2013}. 

\begin{figure}
     \centering
     \begin{subfigure}[b]{\columnwidth}
        \centering
        \includegraphics[width=0.5\columnwidth]{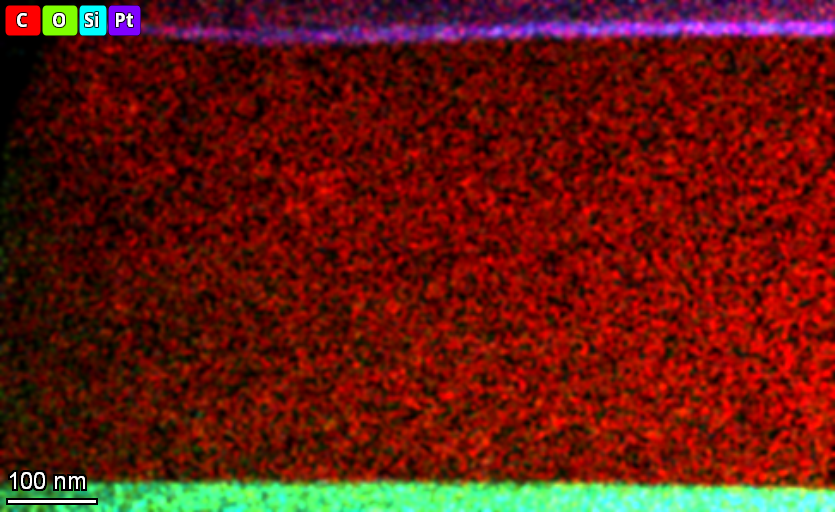}
        \caption{1 mbar water vapor}
         \label{fig:EDS_H2O}
     \end{subfigure}
     \hfill
     \begin{subfigure}[b]{\columnwidth}
        \centering
         \includegraphics[width=0.5\columnwidth]{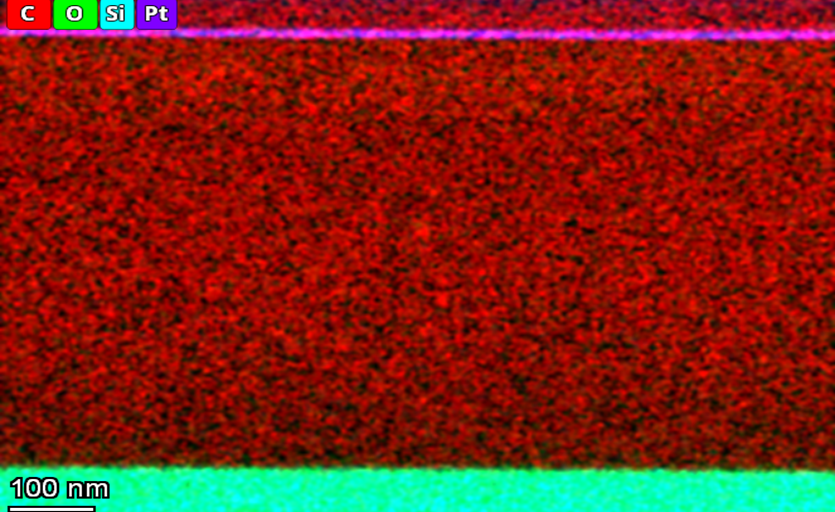}
         \caption{High vacuum}
         \label{fig:EDS_High_Vac}
     \end{subfigure}
        \caption{Elemental mapping EDS image of PS film on soda lime glass substrate irradiated at 15 mC cm$^{-2}$ under (a) 1 mbar water vapor and (b) high vacuum.}
        \label{fig:EDS}
\end{figure}

Elemental mapping EDS image of PS film irradiated at 15 mC/cm$^{2}$ under 1 mbar water vapor (Fig. \ref{fig:EDS_H2O}) and high vacuum (Fig. \ref{fig:EDS_High_Vac}) revealed no signs of film oxidation. This means the origin of PL from the irradiated patterns likely results from the core states of the CD's or PAH's that are formed. The absence of carbonyl groups is also confirmed later in section \ref{FTIR_Glass_all_gases} from the infrared spectrum obtained for the irradiated patterns using FTIR spectroscopy.  

\subsection{PL enhancement with different gases on N-BK7 and soda lime glass substrates: }
In section \ref{N-BK7_vs_SLG}, we demonstrated that the differences in the PL intensity on different substrates with or without the ambient gas in the chamber arises from the differences in the thermal and electrical properties of the substrate. In an environmental scanning electron microscope (ESEM) \cite{danilatos_foundations_1988}, the right selection of the ambient gas, gas pressure, working distance, and electron beam energy allows the imaging of electrically insulating material \cite{toth_effects_2000}. 

Here we expand our choice of ambient gas to include N$_2$, Ar, and He to understand how electron scattering in gases affects the PL intensity. Forward scattering of electrons in gases affects the amount of absorbed dose by the resist and the beam skirt radius from scattering in a gas can be approximated by \cite{danilatos_foundations_1988}

\begin{equation}
    R_s = 364\left( \frac{Z}{E} \right)\left( \frac{P}{T} \right)^{1/2}L^{3/2},    
\end{equation}

where a circle of radius R$_S$, in meters, encompasses 90\% of the scattered electrons, Z is the effective atomic number of the scatterer, E is the primary electron energy in eV, P is the pressure in pascals, T is the temperature in Kelvin, and L is the gas path length in meters. Skirt radius, hence electron scattering in gas, is directly proportional to the effective atomic number of the scatterer gas when the gas pressure, gas path length, temperature and beam energy of the incident electrons are kept constant.

\begin{figure}
    \centering
    \includegraphics[width=0.5\columnwidth]{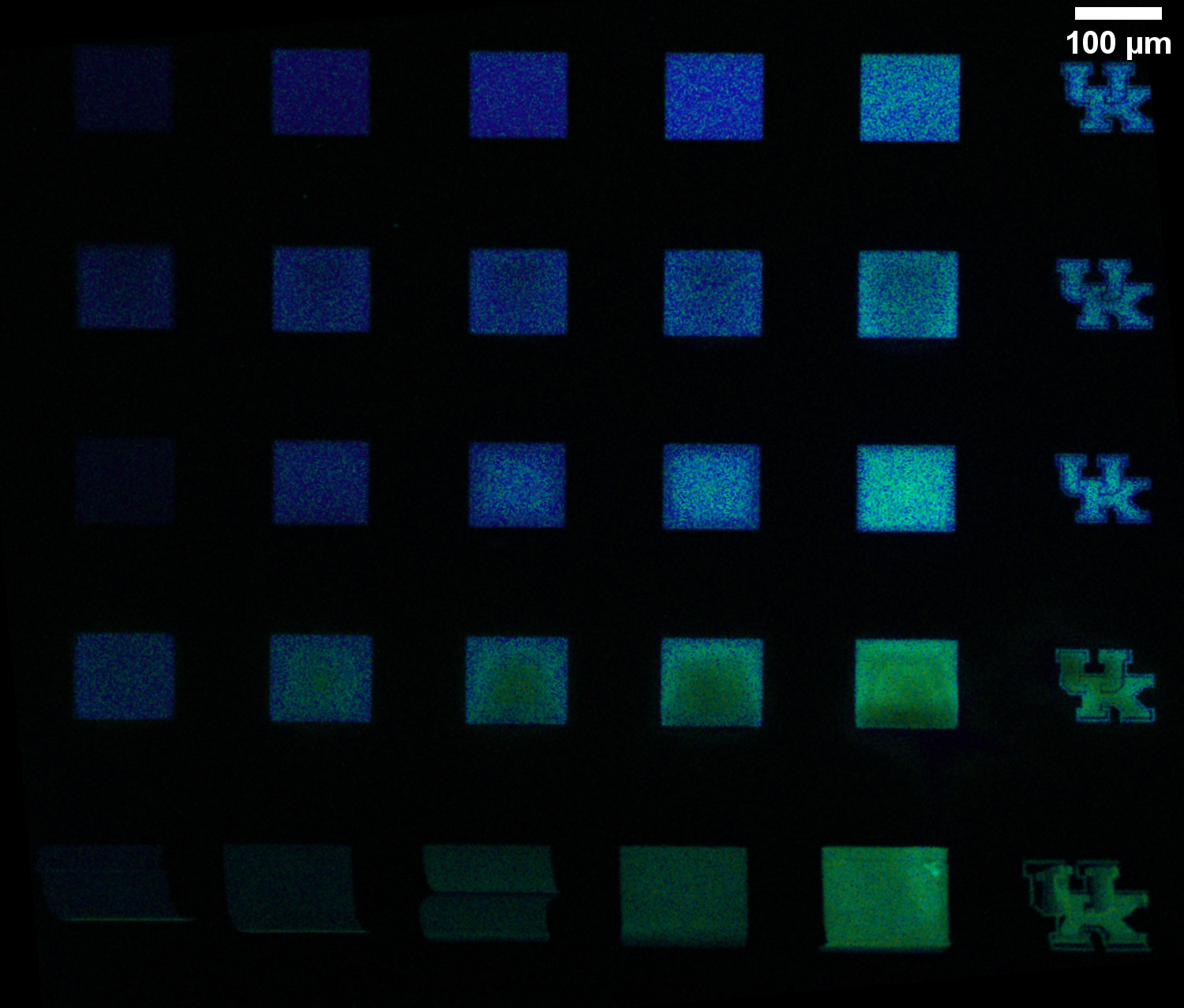}
    \caption{Fluorescence microscopy image of patterns irradiated on soda lime glass substrate. Dose ranged from 1.8 – 15 mC cm$^{-2}$. 100 $\mu$m square patterns were exposed, with a minimum of 100 $\mu$m spacing between each neighboring square, exposed at 950 pA under 1 mbar of gas pressure. The patterns from the top row to the bottom row were exposed under water vapor, nitrogen, argon, helium and high vacuum respectively.}\centering
    \label{fig:Fluorescence_microscopy_SLG_all_gases}
\end{figure}

\begin{figure}
     \centering
     \begin{subfigure}[b]{0.5\columnwidth}
        \centering
        \includegraphics[trim=75 20 90 60,clip,width=\columnwidth]{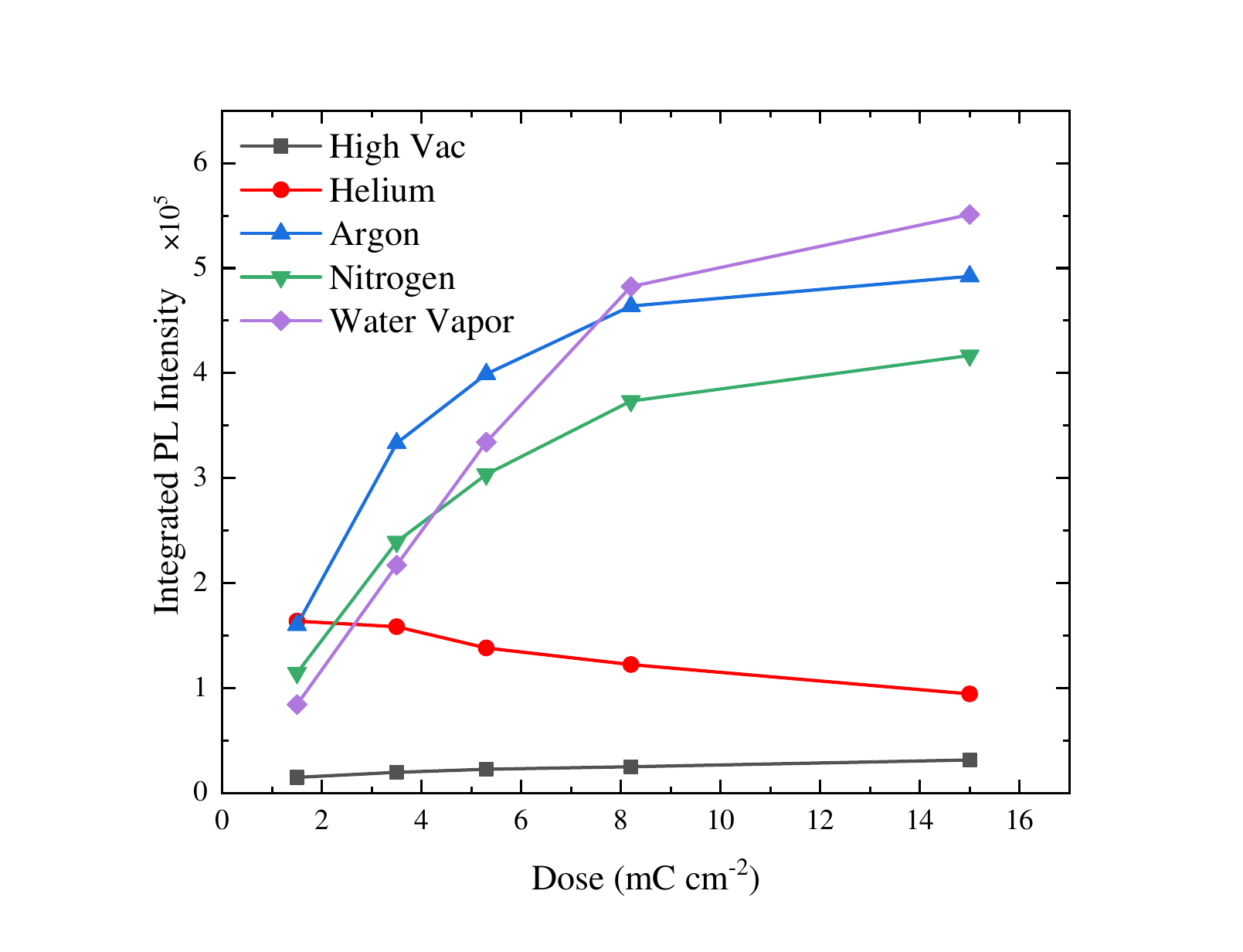}
        \caption{N-BK7}
         \label{fig:IPL_N-BK}
     \end{subfigure}%
     \begin{subfigure}[b]{0.5\columnwidth}
        \centering
         \includegraphics[trim=75 20 90 30,clip,width=\columnwidth]{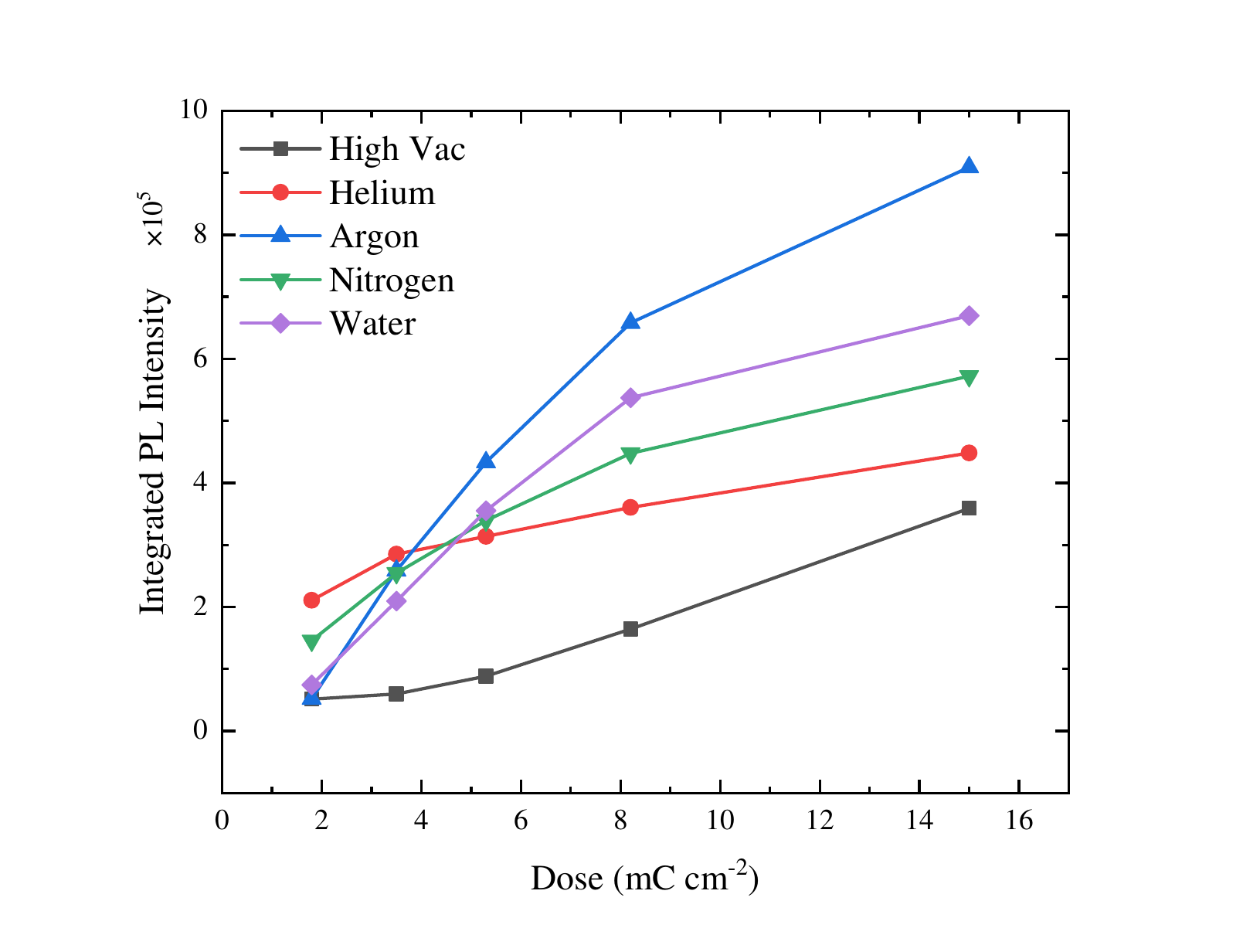}
         \caption{Soda lime glass}
         \label{fig:IPL_Glass}
     \end{subfigure}
        \caption{Integrated PL intensity as a function of electron dose under 1 mbar gas pressure on (a) N-BK7 substrate at a beam current of 764 pA (b) soda lime glass substrate at a beam current of 950 pA.}
        \label{fig:N-BK7_vs_SLG}
\end{figure}

Fluorescence microscopy image of patterns irradiated on soda lime glass substrate under different gases is shown in Figure \ref{fig:Fluorescence_microscopy_SLG_all_gases}. Patterns were exposed at a beam current of 950 pA under 1 mbar of gas pressure. We see that the emission color of the patterns vary with electron dose and ambient gas, no significant scattering is observed for patterns exposed under gaseous environments; for patterns exposed under high vacuum conditions the formation of strong electrostatic fields at the sample surface resulting from substrate charging leads to poor shape fidelity \cite{cummings1989charging}. Comparison of integrated PL intensity for patterns exposed under high vacuum and different gases at 1 mbar as a function of electron dose on N-BK7 (Fig. \ref{fig:IPL_N-BK}) and soda lime glass substrate (Fig. \ref{fig:IPL_Glass}) revealed 

\begin{itemize}
    \item at low doses, gas with lowest scattering cross section, see Table. \ref{gas_scattering_cross_section}, yields the highest PL intensity; whereas at high doses the trend reverses.
    \item integrated PL intensity peaks at lower doses for exposure under helium on N-BK7 substrate.
    \item on N-BK7 substrate exposure under water vapor is more conducive for fluorophore formation while on soda lime glass substrate argon is more effective. 
    \item electron scattering in gas alone is not the dominant mechanism for enhanced PL as trend switches depending on dose and the choice of substrate.
\end{itemize}

It was also found that the emission peak blue-shifts with increasing gas pressure under all gases as  observed for exposure under water vapor and the position of the emission peak is dictated by the gas pressure. The highest PL yield for exposures on N-BK7 and soda lime glass substrate was observed for exposure on soda lime glass substrate under argon gas. Therefore, more extensive characterization were performed only for patterns irradiated on soda lime glass substrate in order to understand how  different gases affects the composition of the irradiated patterns.

$^{1}$1H liquid nuclear magnetic resonance (NMR) and correlation spectroscopy (COSY) has revealed the the existence of several small multi ring aromatic units in irradiated PS films such as biphenyl, p-terphenyl, 1,4-dimethylnaphthalene and 1-phenylnaphthalene; where as phenanthrene was found at higher electron dose irradiation \cite{lee_liquid_2007}. However, these small multi ring aromatic units absorb mostly in the mid UV region and are weakly fluorescent in the visible region with only the tail of the emission spectrum extending in the visible region. Also, the NMR spectra were obtained only for PS samples irradiated at a low exposure dose since cross-linked PS is insoluble in solvents. Therefore, these smaller, multi-ring aromatic species are unlikely to be the source of visible PL observed for higher electron doses.  In fact, the low solubility of irradiated PS and the small absolute number of fluorophores generated by scanning a focused beam limits the range of analysis methods available.  However, micro-Fourier-transform infrared spectroscopy (FTIR) offers one potential path and is discussed next. 

\subsubsection{Fourier-transform infrared spectroscopy for patterns exposed on soda lime glass substrates: }
\label{FTIR_Glass_all_gases}

\begin{table}
\centering
\caption{\label{gas_scattering_cross_section}Total gas scattering cross-section at 20 keV}
\begin{tabular}{@{}|l|c|}
\hline
Gas&Total gas scattering cross-section (cm$^2$)\\ \hline
Argon&2.14 x 10$^{-17}$ \cite{he_measurement_2003}\\ \hline
Nitrogen&1.62 x 10$^{-17}$ \cite{rattenberger_method_2009}\\ \hline
Water&1.18 x 10$^{-17}$ \cite{rattenberger_method_2009}\\ \hline
Helium&1.73 x 10$^{-18}$ \cite{he_measurement_2003}\\ \hline
\end{tabular}\\
\end{table}

Important vibration bands in PS, listed in Table \ref{FTIR_peak_assignments}, lie at: 1453 $cm^{-1}$ for the —CH$_2$ bend of the carbon backbone; 1602 $cm^{-1}$ for aromatic sp$^2$ carbon stretches; 2851 and 2925 $cm^{-1}$ for the symmetric and asymmetric C—H stretches on the backbone; and 1492, 3029, 3061 and 3083 $cm^{-1}$ for the aromatic C—H stretches.

\begin{figure*}
     \centering
     \begin{subfigure}[b]{0.45\textwidth}
         \centering
         \includegraphics[width=\textwidth]{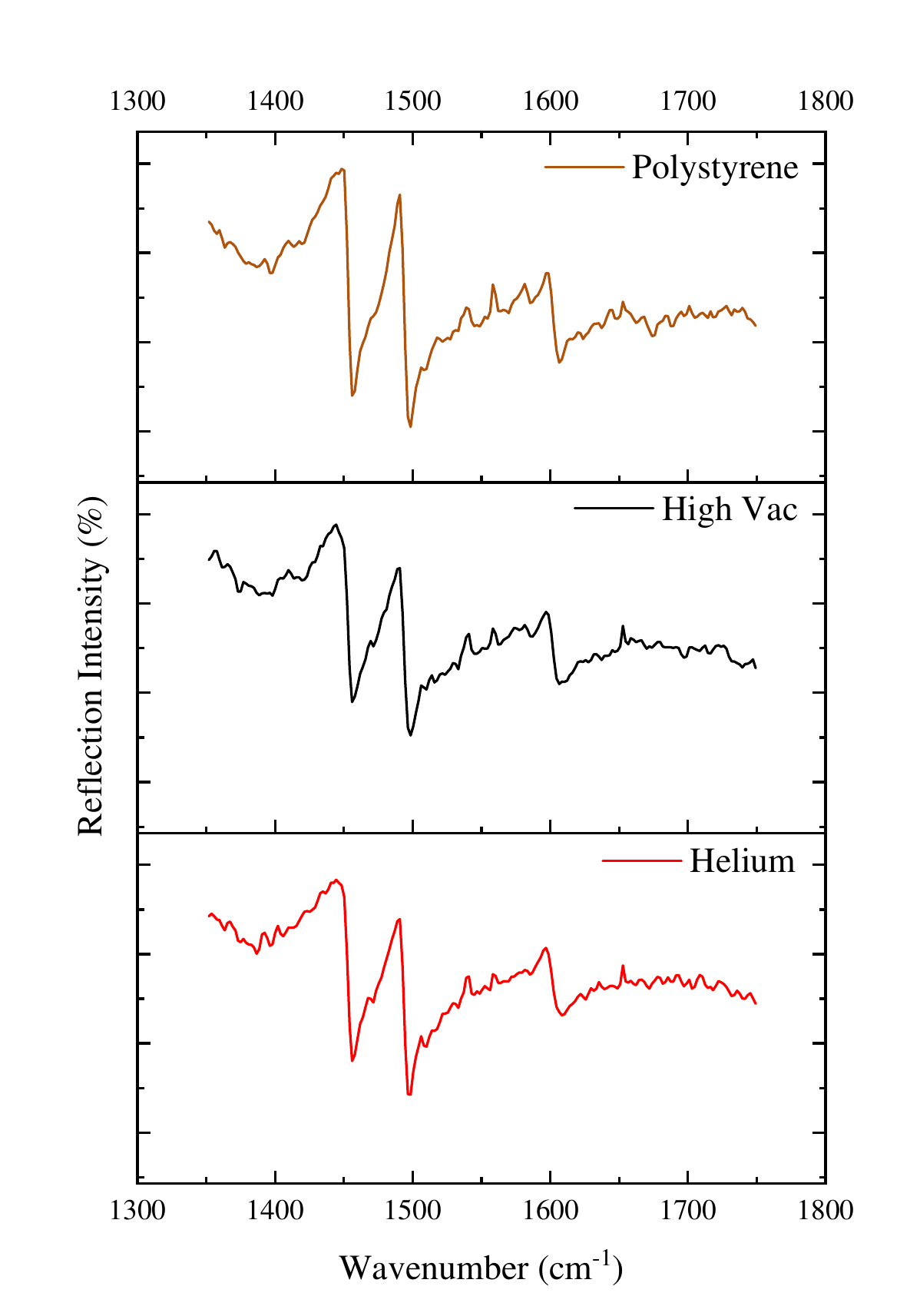}
         \caption{PS, High Vacuum and Helium}
         \label{fig:FTIR_PS_HV_He_1.8mC_1350-1750}
     \end{subfigure}
     \hfill
     \begin{subfigure}[b]{0.45\textwidth}
        \centering
         \includegraphics[width=\textwidth]{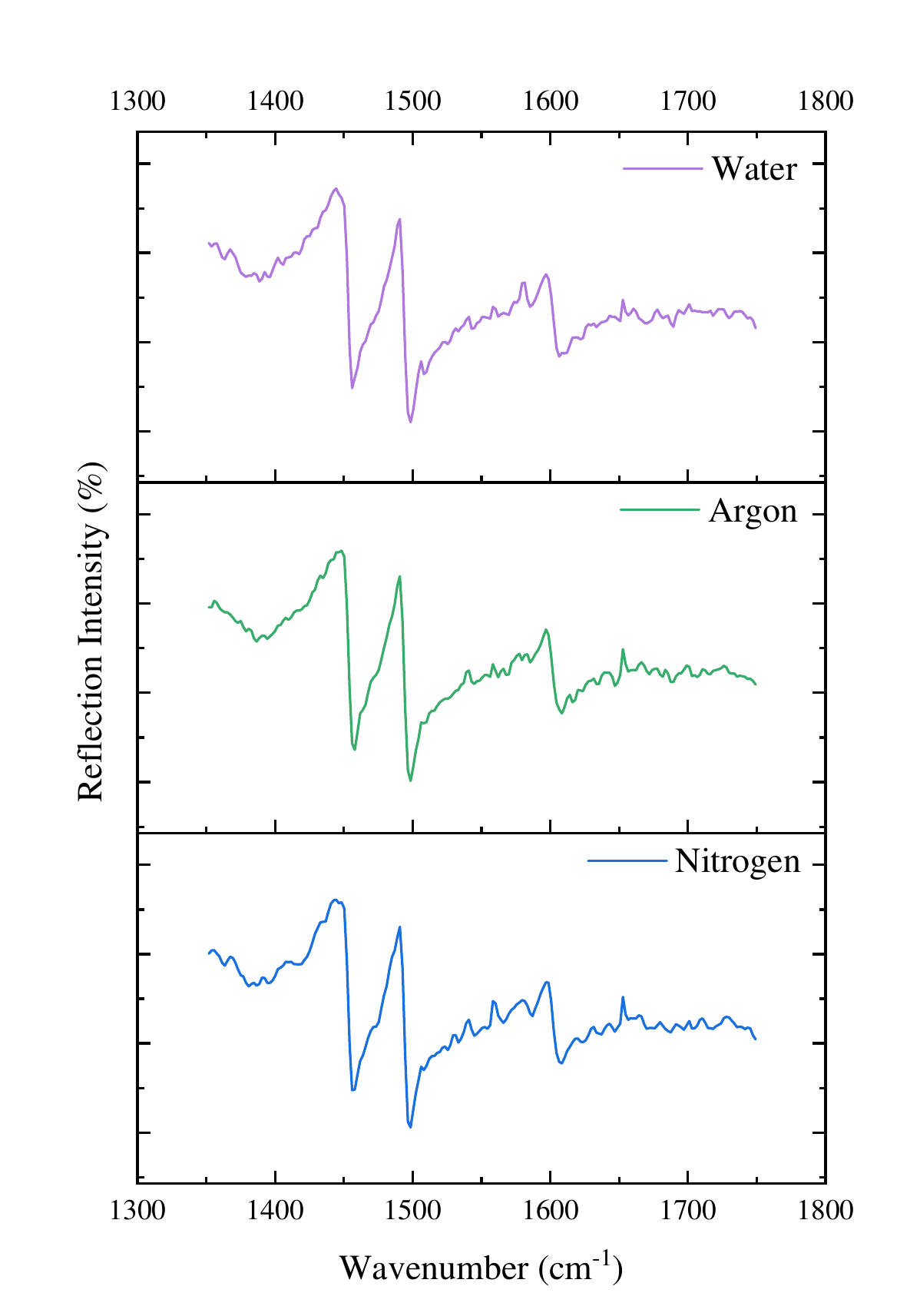}
         \caption{Water, Argon and Nitrogen}
         \label{fig:FTIR_W_A_N_1.8mC_1350-1750}
     \end{subfigure}
     \hfill
     \centering
     \begin{subfigure}[b]{0.45\textwidth}
         \centering
         \includegraphics[width=\textwidth]{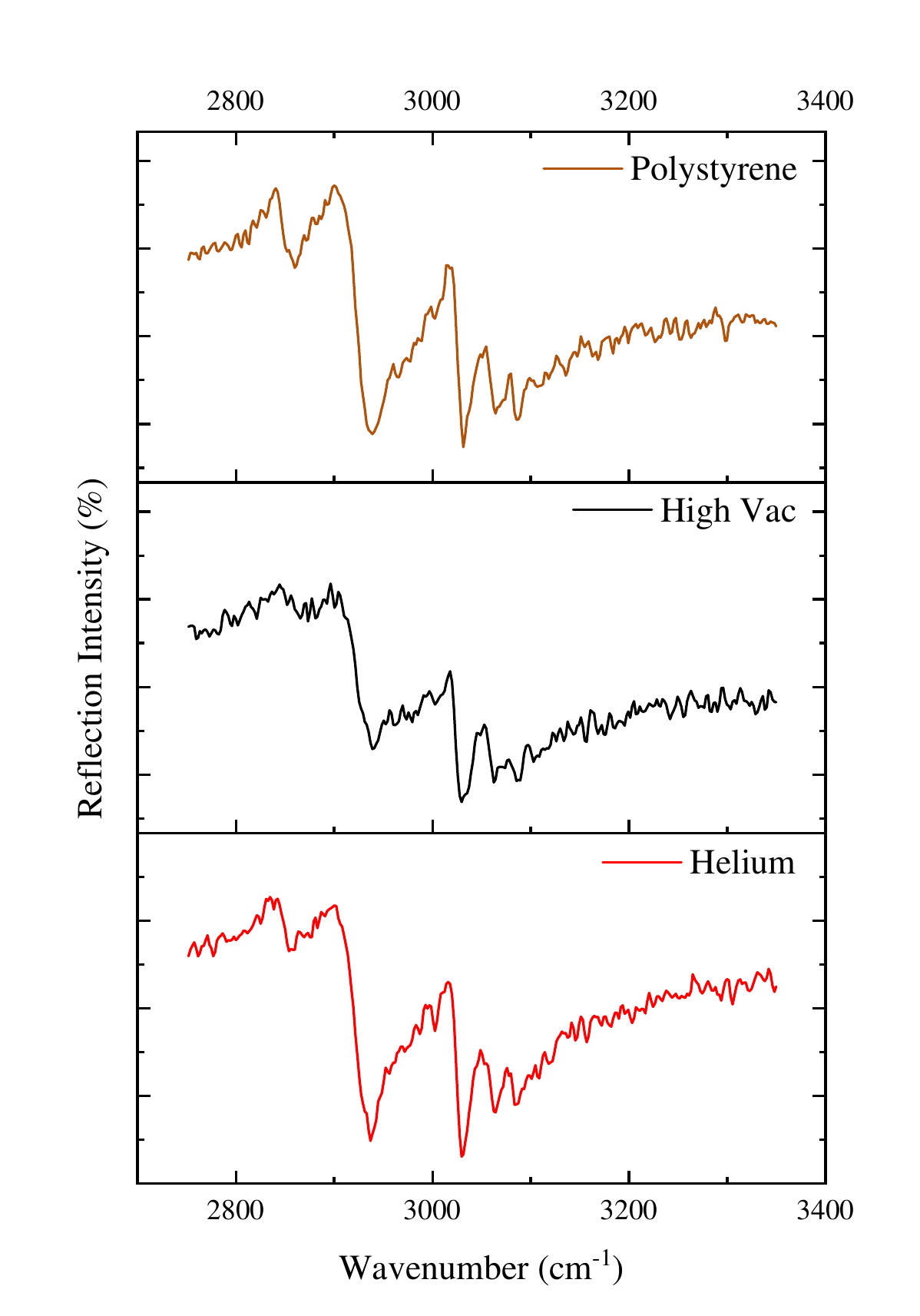}
         \caption{PS, High Vacuum and Helium}
         \label{fig:FTIR_PS_HV_He_1.8mC_2750-3350}
     \end{subfigure}
     \hfill
     \begin{subfigure}[b]{0.45\textwidth}
        \centering
         \includegraphics[width=\textwidth]{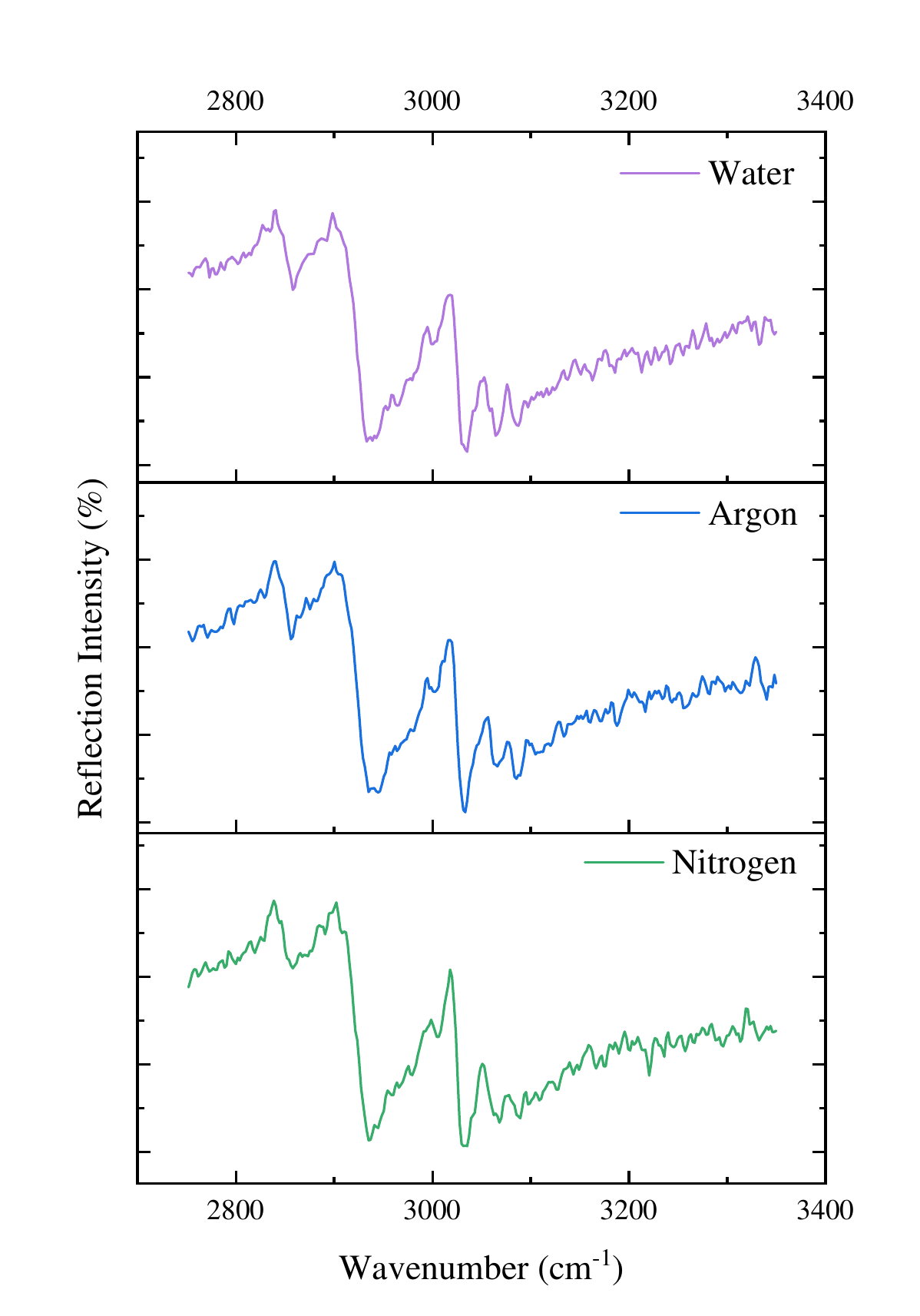}
         \caption{Water, Argon and Nitrogen}
         \label{fig:FTIR_W_A_N_1.8mC_2750-3350}
     \end{subfigure}
        \caption{Exposure under high vacuum and 1 mbar gas pressure at 1.8 mC cm$^{-2}$}
        \label{fig:FTIR_on_Glass_all_gasses_1.8mC}
\end{figure*}

\begin{figure*}
     \centering
     \begin{subfigure}[b]{0.45\textwidth}
         \centering
         \includegraphics[width=\textwidth]{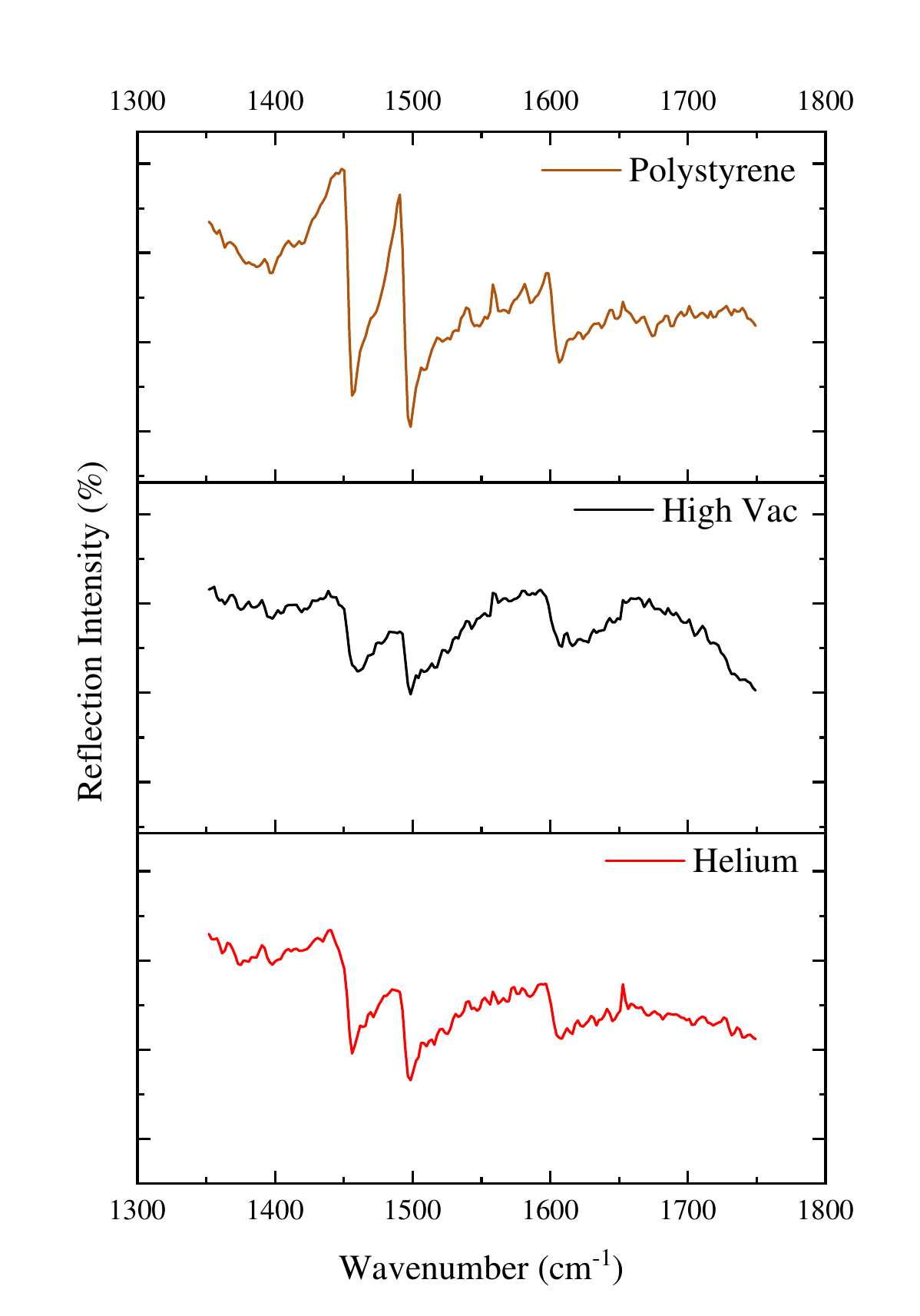}
         \caption{PS, High Vacuum and Helium}
         \label{fig:FTIR_PS_HV_He_15mC_1350-1750x}
     \end{subfigure}
     \hfill
     \begin{subfigure}[b]{0.45\textwidth}
        \centering
         \includegraphics[width=\textwidth]{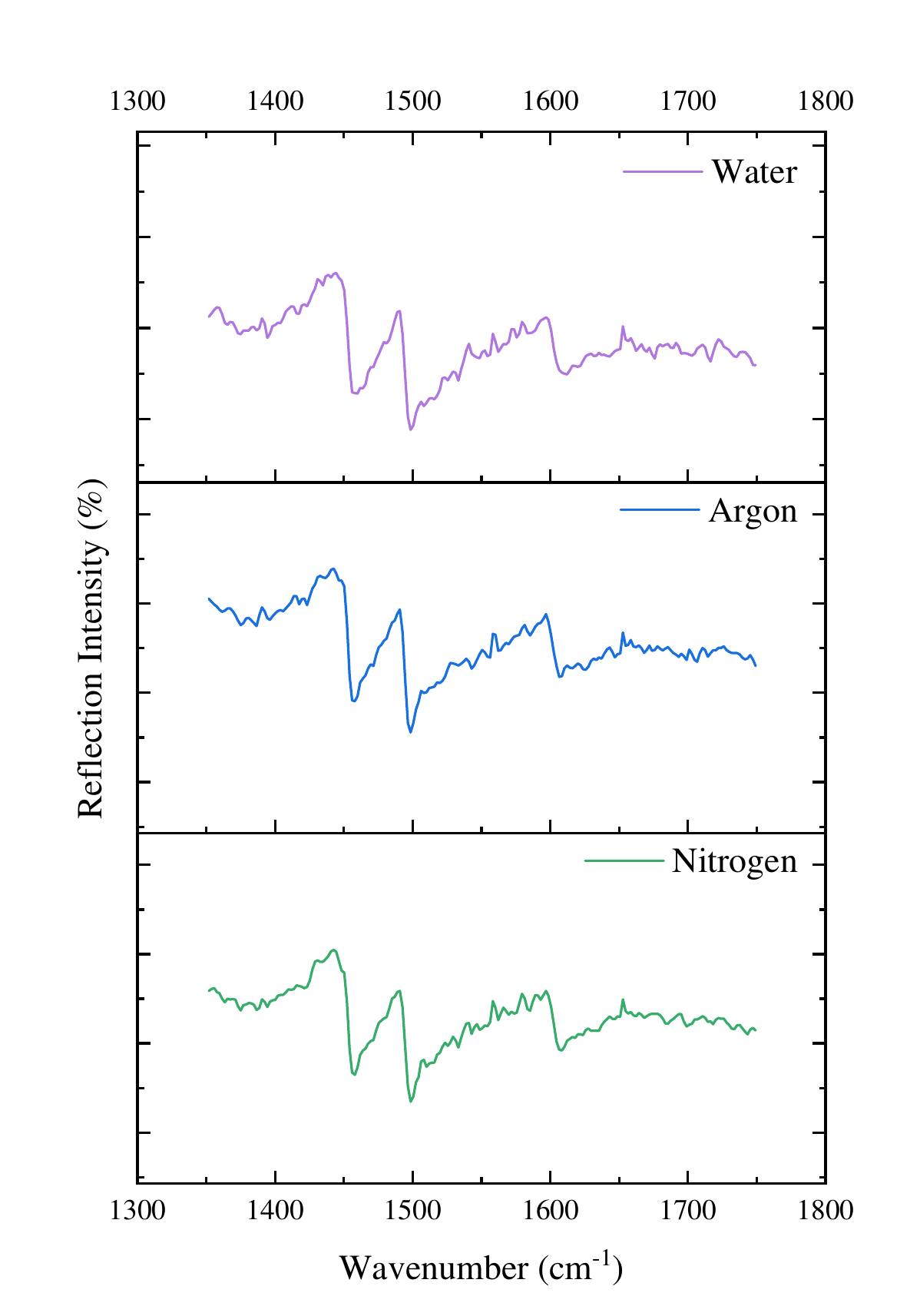}
         \caption{Water, Argon and Nitrogen}
         \label{fig:FTIR_W_A_N_15mC_1350-1750}
     \end{subfigure}
     \hfill
     \centering
     \begin{subfigure}[b]{0.45\textwidth}
         \centering
         \includegraphics[width=\textwidth]{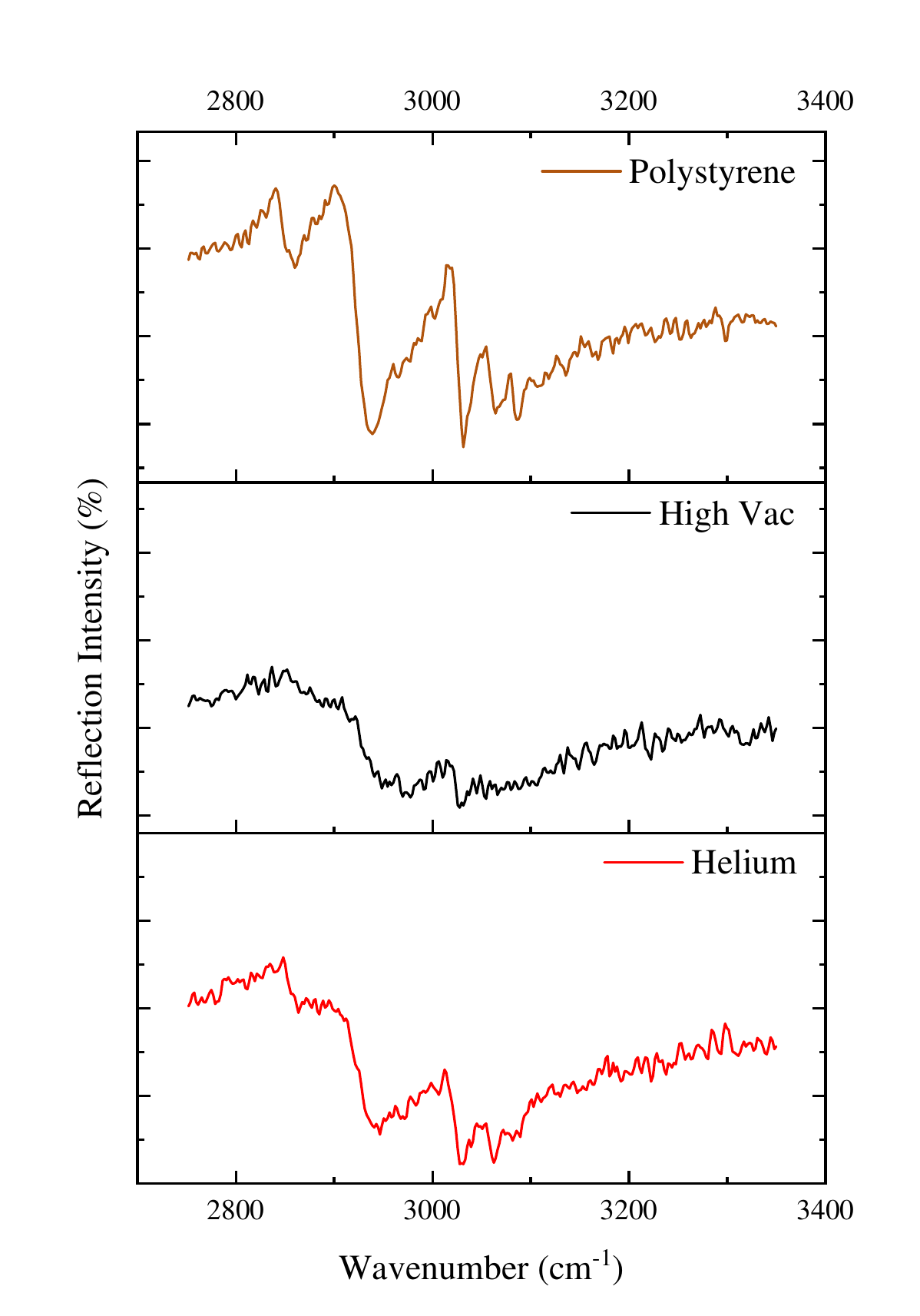}
         \caption{PS, High Vacuum and Helium}
         \label{fig:FTIR_PS_HV_He_15mC_2750-3350}
     \end{subfigure}
     \hfill
     \begin{subfigure}[b]{0.45\textwidth}
        \centering
         \includegraphics[width=\textwidth]{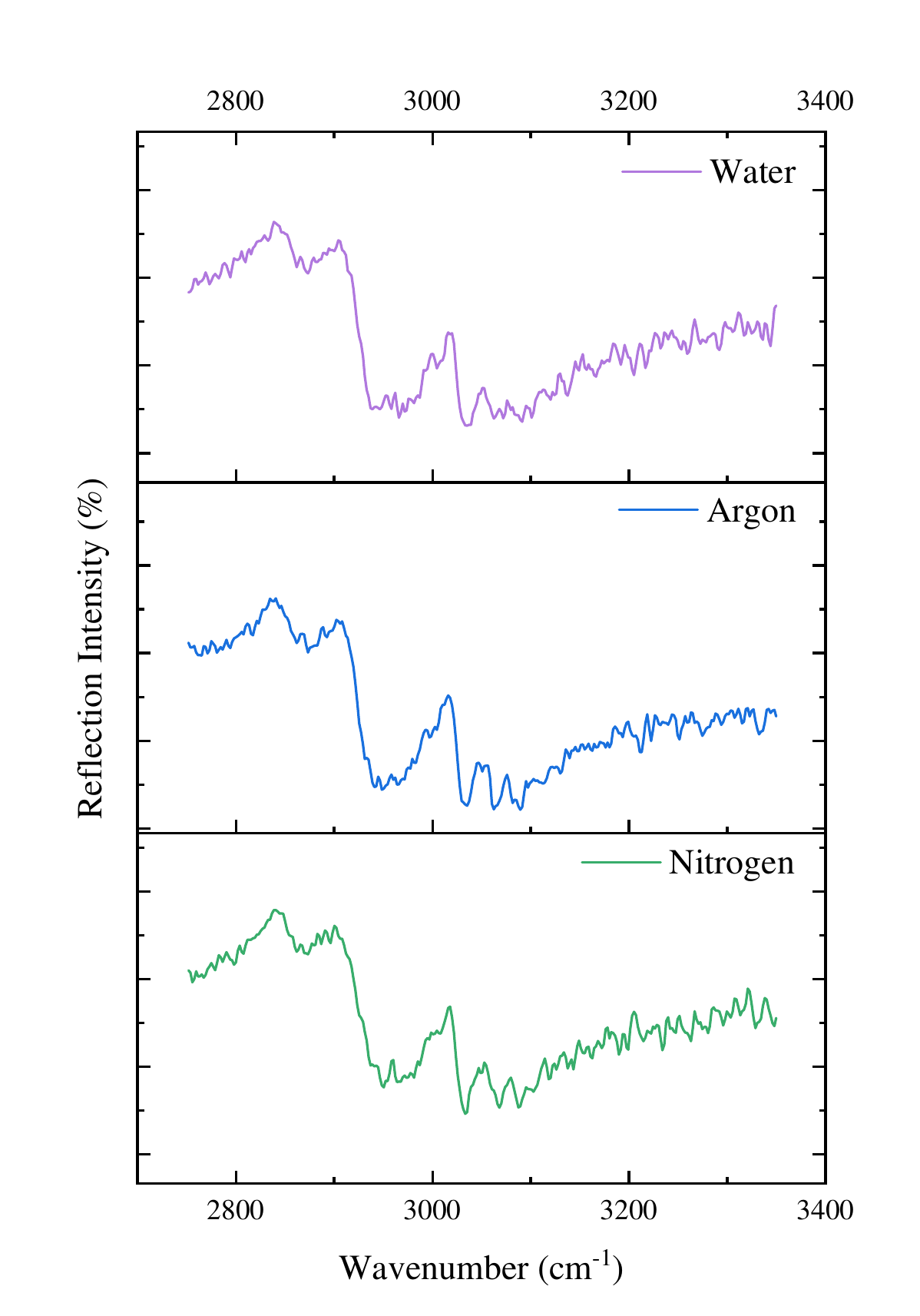}
         \caption{Water, Argon and Nitrogen}
         \label{fig:FTIR_W_A_N_15mC_2750-3350}
     \end{subfigure}
        \caption{Exposure under high vacuum and 1 mbar gas pressure at 15 mC cm$^{-2}$}
        \label{fig:FTIR_on_Glass_all_gasses_15mC}
\end{figure*}

Under high vacuum as well as for exposure under gases, Fig. \ref{fig:FTIR_on_Glass_all_gasses_1.8mC} and \ref{fig:FTIR_on_Glass_all_gasses_15mC}, the aromatic sp$^2$ carbon are preserved which suggests the PL observed upon irradiation originates from the formation of polycyclic structures. Under high vacuum the —CH$_2$ of the backbone, the C—H of the backbone, and the C-H of the aromatic ring dissociate more rapidly compared to exposure under gases. The introduction of gases appears to preserve hydrogen both on the backbone and the aromatic ring as well as slow down the breaking of C—C backbone. Thus, under gaseous environments the decay of aromatic and aliphatic C–H stretches is reduced compared to high vacuum exposure; in all cases (under high vacuum as well as gases), features associated with the phenyl rings are preserved, resulting in enhanced PL from PS irradiated under gases. 

\begin{table}
\centering
\caption{Important vibrational mode assignments for PS from FTIR reflection data}
\label{FTIR_peak_assignments}
\begin{tabular}{@{}|l|l|l|}
\hline
\multicolumn{1}{|c}{\textbf{Wavenumber (cm$^{-1}$)}}& \multicolumn{2}{|c|}{\textbf{Modes}}\\
\hline
1453& —CH${_2}$&bending\\ \hline
1492&  C—H& aromatic stretches\\ \hline
1602&  C=C& aromatic stretches\\ \hline
2851 and 2925&  C—H& symmetric \& asymmetric stretches (backbone)\\ \hline
3029, 3061 and 3083&  C—H& aromatic stretches\\ \hline
\end{tabular}\\
\end{table}

\subsection{PL enhancement on other insulating, conducting and semi-conducting substrates under ambient gases:}
\label{Insulating_vs_conducting_substrates}

The presence of a phenyl group in a material's molecular structure also has a significant impact on its charging properties. The phenyl rings act as deep traps to prevent redistribution of electrons which have lost their kinetic energy and stopped in the PS.  It has also been proposed that higher kinetic-energy electrons can trapped by the phenyl rings, where their kinetic energy is lost by the resonance effect. As a result, the chain scission of the polymer backbone is prevented.\cite{nagasawa_charge_2008, nagasawa_charge_2010} 

\begin{figure}
     \centering
     \begin{subfigure}[b]{0.5\columnwidth}
         \centering
         \includegraphics[trim=75 20 90 60,clip,width=\columnwidth]{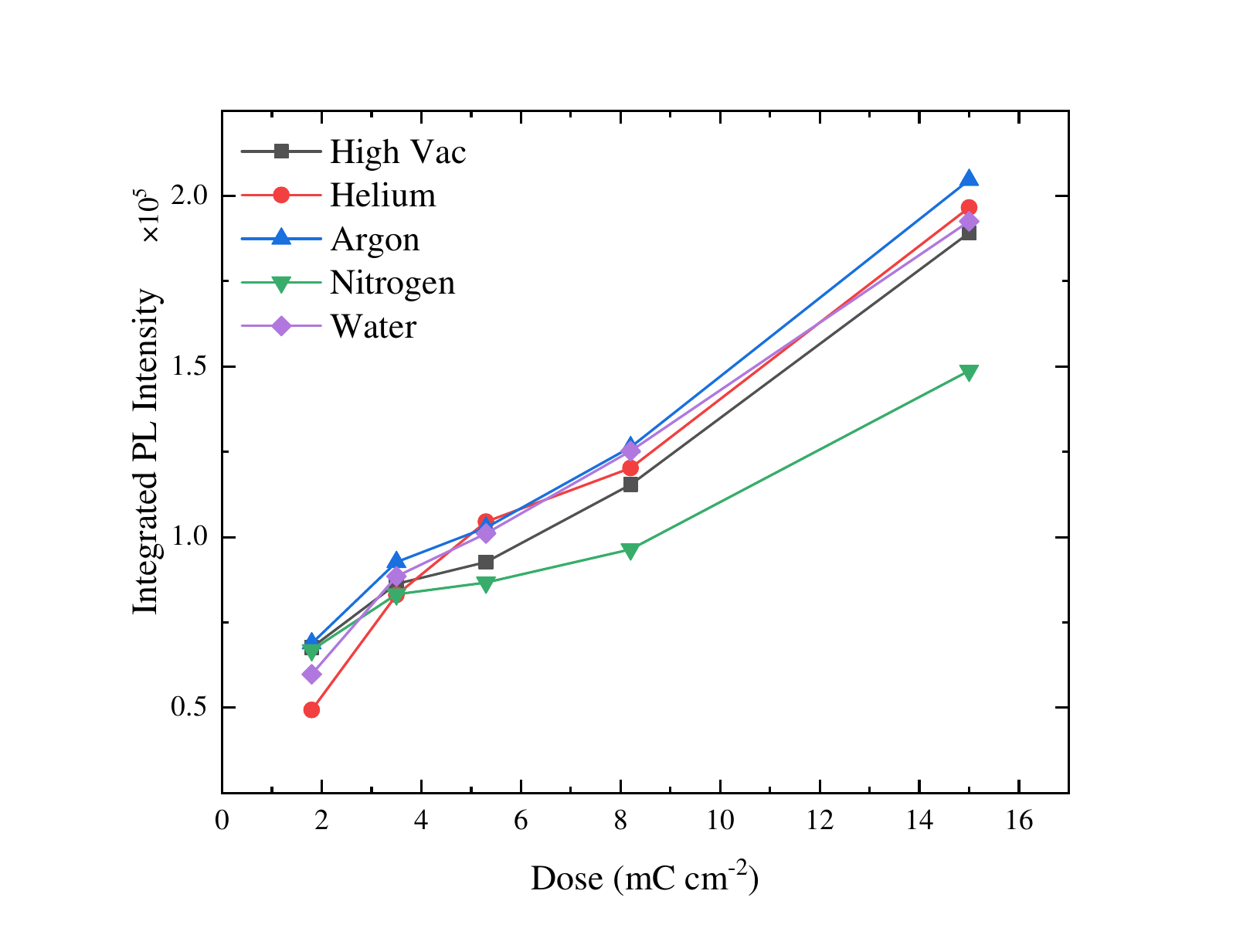}
         \caption{Silicon}
         \label{fig:Int_PL_Si}
     \end{subfigure}%
     \begin{subfigure}[b]{0.5\columnwidth}
        \centering
         \includegraphics[trim=75 20 90 30,clip,width=\columnwidth]{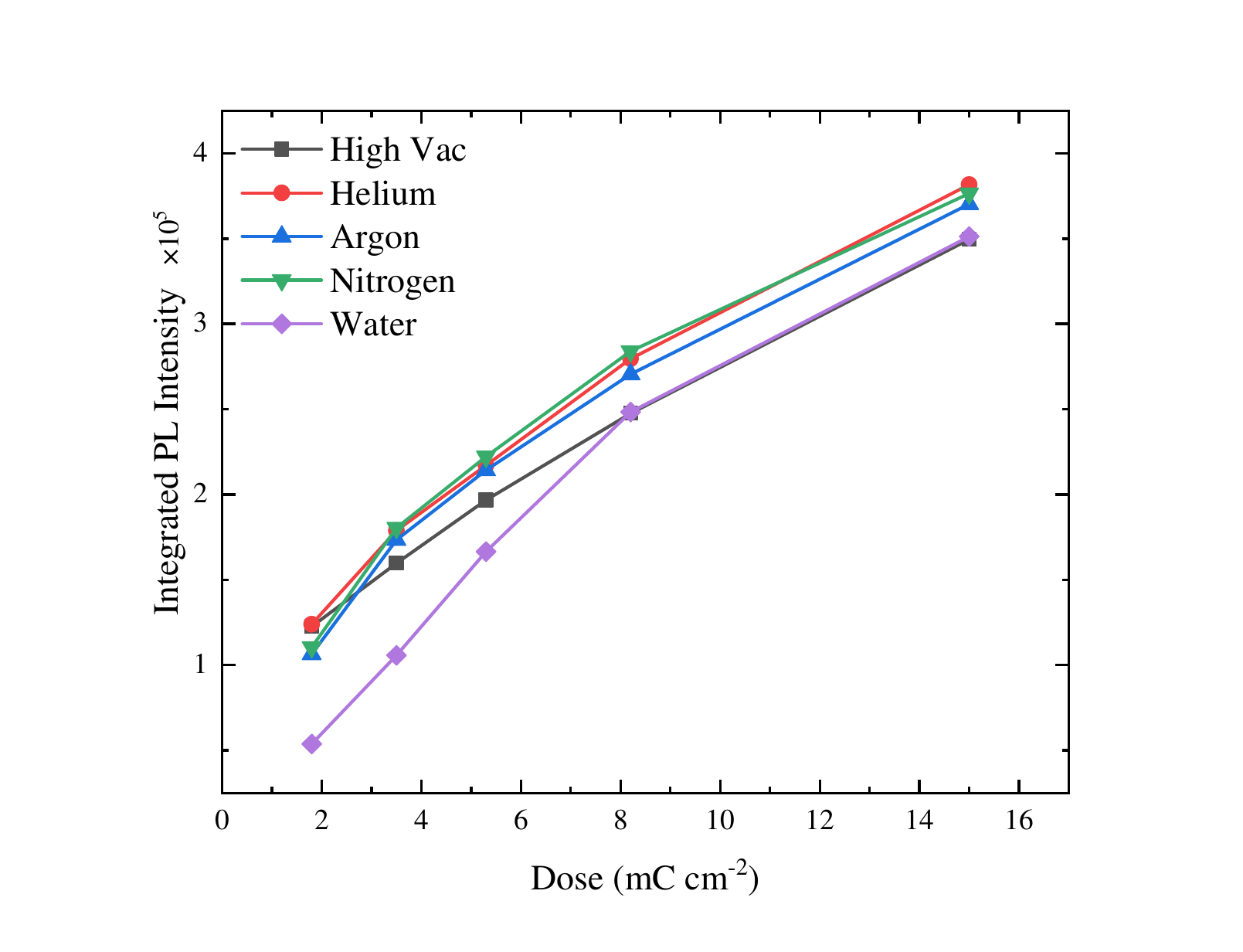}
         \caption{ITO coated soda lime glass}
         \label{fig:Int_PL_ITO}
     \end{subfigure}
     \caption{Integrated PL intensity on conducting substrates as a function of electron dose under 1 mbar gas pressure exposed at a beam current of 950 pA on (a) Silicon and (b) ITO coated soda lime glass.}
        \label{fig:Int_PL_Conduting_Substrates}
\end{figure}

\begin{figure}
     \centering
     \begin{subfigure}[b]{0.5\columnwidth}
         \centering
         \includegraphics[trim=75 20 90 60,clip,width=\columnwidth]{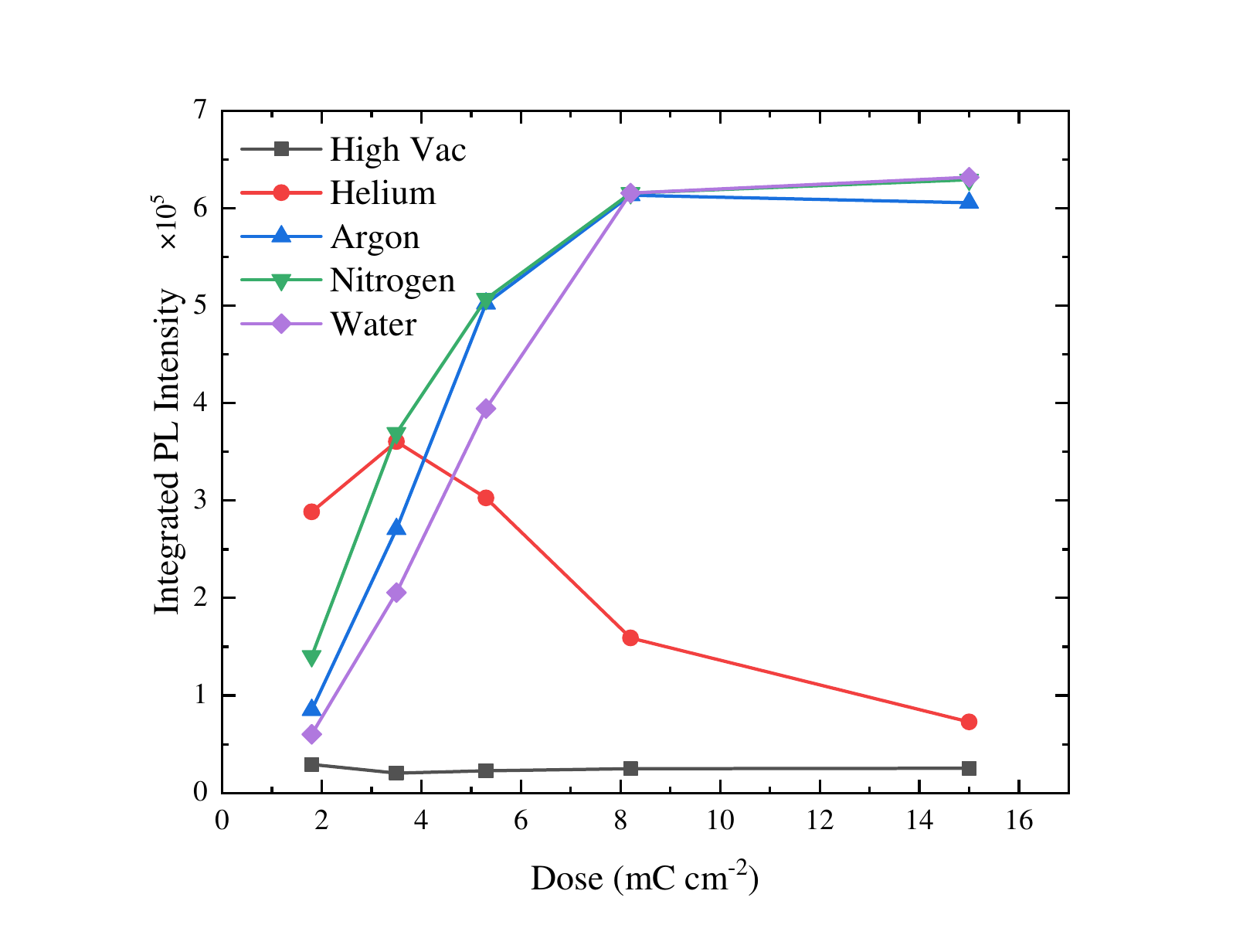}
         \caption{Fused Silica}
         \label{fig:Int_PL_FS}
     \end{subfigure}%
     \begin{subfigure}[b]{0.5\columnwidth}
        \centering
         \includegraphics[trim=75 20 90 30,clip,width=\columnwidth]{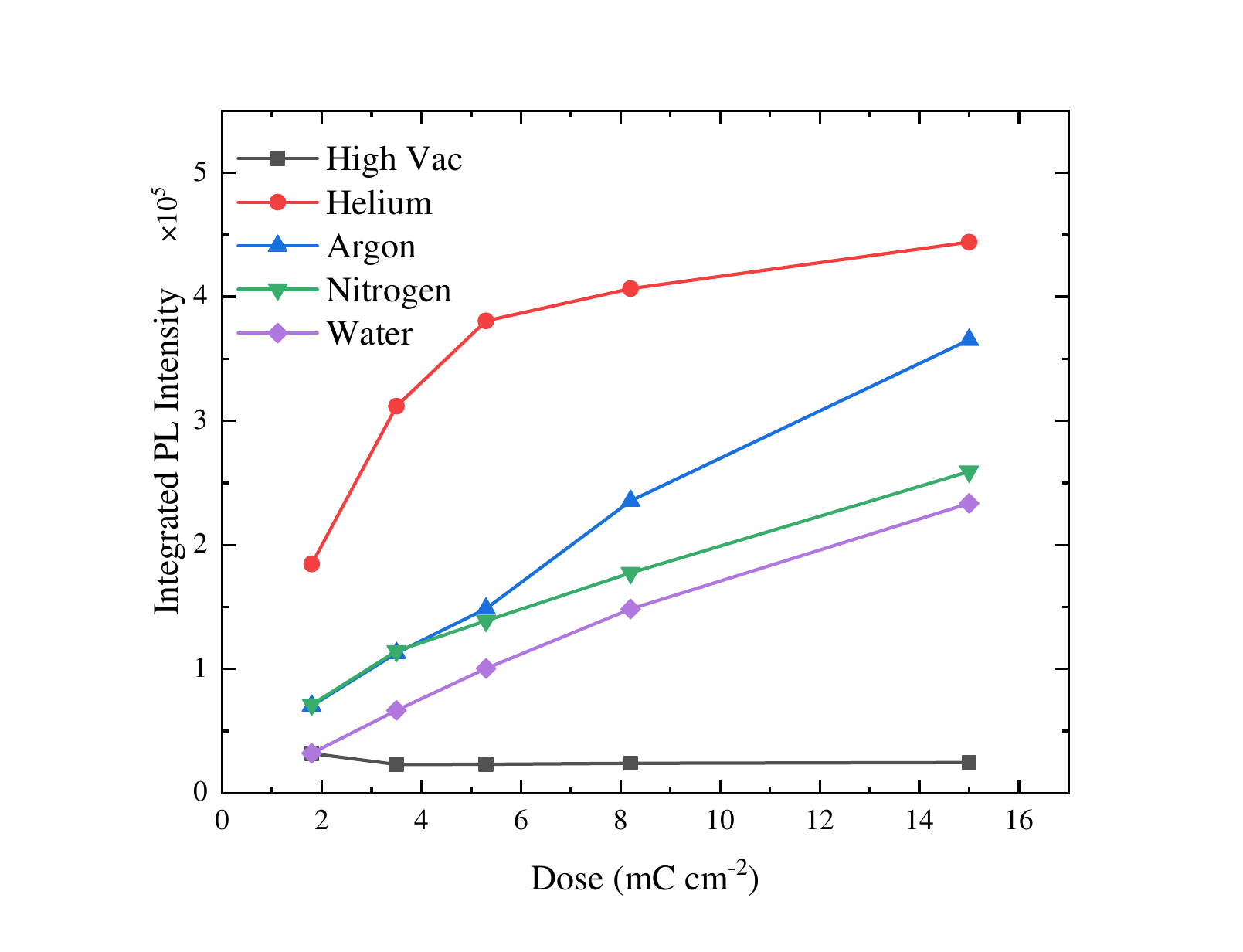}
         \caption{Sapphire}
         \label{fig:Int_PL_Sapphire}
     \end{subfigure}
        \caption{Integrated PL intensity on instulating substrates as a function of electron dose under 1 mbar gas pressure exposed at a beam current of 950 pA on (a) Fused Silica and (b) Sapphire.}
        \label{fig:Int_PL_Insulating_Substrates}
\end{figure}

Silicon and ITO coated glass dissipate charge and the integrated PL intensity is proportional to the electron dose and is almost independent of the gas (Fig. \ref{fig:Int_PL_Conduting_Substrates}). Thus, on conductive substrates PL is not influenced by the presence of a gas. Photon-yield on conductive substrates is significantly smaller than that yield from insulating substrates. On sapphire the integrated PL intensity is factor of 2 or more higher than on Si (sapphire and silicon have similar thermal conductivity, see Table \ref{substrate_thermal_conductivity}), and PL shows a gas dependence. Integrated PL intensity varies inversely with the thermal conductivity of substrates; soda lime glass being least thermally conductive yields the highest PL yield.  This suggests that e-beam heating plays some role in the fluorophore formation process.

On electrically insulating substrates the PL is influenced by the presence of a gas (Fig. \ref{fig:Int_PL_Insulating_Substrates}), this could possibly result from the differences in the charge dissipation with different gases. The degree of charge dissipation depends on the ionization coefficient of gases as well as the experimental conditions used such as working distance and the gas pressure. With fixed working distance and gas pressure the degree of charge dissipation is a function of ionization coefficient of gases. Gases mitigate charging to different degrees and contribute to the enhancement in PL. Thus, on substrates with similar thermal conductivity, substrate charging affects the process of fluorophore formation and significantly influences the PL.

\begin{table}
\centering
\caption{\label{substrate_thermal_conductivity}Thermal conductivity of substrates at room temperature}
\begin{tabular}{@{}|l|c|}
\hline
Substrate&Thermal conductivity (Wm$^{-1}$K$^{-1}$)\\ \hline
Soda lime glass&7.31 x 10$^{-1}$ \cite{ito}\\ \hline
N-BK7&1.05 x 10$^0$ \cite{a2024_materials}\\ \hline
Fused Silica&1.38 x 10$^0$ \cite{a2024_materials}\\ \hline
ITO&4.00 x 10$^0$ \cite{ashida2009thermal}\\ \hline
Sapphire&4.39 x 10$^1$ \cite{a2024_materials}\\ \hline
Silicon&1.26 x 10$^2$ \cite{a2024_materials}\\ \hline
\end{tabular}\\
\end{table}

\begin{table}
\centering
\caption{\label{gross_ionization_cross_section}Gross ionization cross-section at 600 eV}
\begin{tabular}{@{}|l|c|}
\hline
Gas&Gross ionization cross-section (cm$^2$)\\ \hline
Argon&1.20 x 10$^{-16}$ \cite{schram_ionization_1965}\\ \hline
Nitrogen&1.15 x 10$^{-16}$ \cite{schram_ionization_1965}\\ \hline
Water&1.00 x 10$^{-16}$ \cite{schutten_gross_1966}\\ \hline
Helium&1.59 x 10$^{-17}$ \cite{schram_ionization_1965}\\ \hline
\end{tabular}\\
\end{table}

Gas influences the charging of substrates affecting the formation of fluorophores resulting in an enhanced integrated PL intensity on soda lime glass where the 5x enhancement compared to Si is seen. Highest integrated PL intensity is observed on soda lime glass substrate, which is least thermally conductive of the substrates used in this study. As we have seen earlier electron scattering in gas does not significantly affect the PL but slight differences in thermal conductivity's of substrates results in different PL characteristics \ref{N-BK7_vs_SLG}, the difference in integrated PL intensity on soda lime glass and Si can be attributed to the differences in thermal conductivity of the substrates. 
Argon and water have markedly different ionization cross-sections (Table. \ref{gross_ionization_cross_section}) which results in the differences in charging and subsequently to the differences in PL. Thus, substrate charging, influenced by both the substrate material and the gas, and thermal conductivity of substrate influences the process of fluorophore formation, significantly influencing the PL.

\section{Summary and Conclusions}

The introduction of gases provides a novel way to tune the emission wavelength and enhance the fluorescence intensity of irradiated polystyrene.  PL intensity of patterns exposed under high vacuum indicates a red-shift in the emission wavelength and an increase in the PL intensity as the electron dose is increased.  This is consistent with prior work on electron irradiated PS. The novelty of this work is to tune the emission wavelength and photon yield simultaneously by varying both dose and gas pressure. For exposure under gaseous environment, the integrated PL increases with dose and then gradually decreases after a critical electron dose is reached. 

Water vapor significantly increased the photon yield on N-BK7 substrates with a maximum occurring at 1 mbar pressure, a result which cannot be explained by simple electron scattering in the gas.  The emission peak blue-shifts with increasing gas pressure and can be tuned over a wide wavelength range. Comparison of integrated PL intensity on N-BK7 and soda lime glass substrates revealed that differences in electrical and thermal properties of substrates leads to different PL characteristics for exposures under otherwise identical conditions.

On N-BK7 substrates, exposure under water vapor is more conducive for fluorophore formation while on soda lime glass substrates argon is more effective. For exposure on soda lime glass substrates under water vapor, high-resolution transmission electron microscope images revealed e-beam irradiated PS is amorphous in nature and elemental mapping EDS revealed no signs of oxidation of the film. FTIR spectroscopy revealed that under gaseous environment the decay of aromatic and aliphatic C–H stretches is reduced compared to the high vacuum exposure; in all cases, features associated with the phenyl rings are preserved. 

Up to 18x enhancement in fluorescence yield can be achieved on sapphire substrates for exposure under helium. Overall, the highest PL yield is observed on soda lime glass substrate under argon environment. Photon-yield on conductive substrates is significantly smaller than that yield from insulating substrates. Substrate charging, thermal conductivity of substrates as well as the gas species influences the process of fluorophore formation, significantly influencing the PL. The differences in the PL characteristics could arise from factors such as higher thermal conductivity of or fluorescence quenching by conductive substrates. 

While understanding all of the interacting physical mechanisms for wavelength tuning and PL enhancement is far from complete, we can conclude that 
\begin{itemize}
    \item Electron scattering in the gas is not responsible for the enhanced PL.
    \item Electrically insulating substrates exhibit gas- and beam-current dependent PL suggesting that charge mitigation plays an important role.
    \item PL is further affected by the thermal conductivity of the substrate, suggesting that e-beam heating plays an important role.
\end{itemize}

Thus, localized e-beam synthesis of fluorophores in PS can be controlled by both dose and by ambient gas pressure. This technique could enable new approaches to photonics where fluorophores with tunable emission properties can be locally introduced by e-beam patterning.

\section{Acknowledgments}

This work was supported by the National Science Foundation (NSF) under Grant No. CMMI-2135666. This work was performed, in part, at the University of Kentucky Center for Nanoscale Science and Engineering and Electron Microscopy Center, members of the National Nanotechnology Coordinated Infrastructure (NNCI), which was supported by the National Science Foundation (No. NNCI-2025075).

\bibliographystyle{unsrt}
\bibliography{reference} 

\begin{thebibliography}{10}

\bibitem{lai_experimental_1979}
Juey~H. Lai and Lloyd~T. Shepherd.
\newblock Experimental {Observations} of {Nearly} {Monodisperse} {Polystyrene} as {Negative} {Electron} {Resists}.
\newblock {\em Journal of The Electrochemical Society}, 126(4):696, April 1979.

\bibitem{itaya_high_1982}
Kingo Itaya, Kimio Shibayama, and Teruo Fujimoto.
\newblock High {Resolution} {Electron} {Beam} {Negative} {Resist} with {Very} {Narrow} {Molecular} {Weight} {Distributions}.
\newblock {\em Journal of The Electrochemical Society}, 129(3):663, March 1982.

\bibitem{manako_nanometer-scale_1997}
Shoko Manako, Jun-ichi Fujita, Yukinori Ochiai, Eiichi~Nomura Eiichi~Nomura, and Shinji~Matsui Shinji~Matsui.
\newblock Nanometer-{Scale} {Patterning} of {Polystyrene} {Resists} in {Low}-{Voltage} {Electron} {Beam} {Lithography}.
\newblock {\em Japanese Journal of Applied Physics}, 36(12S):7773, December 1997.

\bibitem{manako_resolution-limit_1997}
Shoko Manako, Jun-ichi Fujita, Yukinori Ochiai, Eiichi~Nomura Eiichi~Nomura, and Shinji~Matsui Shinji~Matsui.
\newblock Resolution-{Limit} {Study} of {Chain}-{Structure} {Negative} {Resist} by {Electron} {Beam} {Lithography}.
\newblock {\em Japanese Journal of Applied Physics}, 36(6A):L724, June 1997.

\bibitem{ma_polystyrene_2011}
Siqi Ma, Celal Con, Mustafa Yavuz, and Bo~Cui.
\newblock Polystyrene negative resist for high-resolution electron beam lithography.
\newblock {\em Nanoscale Research Letters}, 6(1):446, July 2011.

\bibitem{con_high_2012}
Celal Con, Ripon Dey, Mark Ferguson, Jian Zhang, Raafat Mansour, Mustafa Yavuz, and Bo~Cui.
\newblock High molecular weight polystyrene as very sensitive electron beam resist.
\newblock {\em Microelectronic Engineering}, 98:254--257, October 2012.

\bibitem{con_dry_2013}
Celal Con, Arwa~Saud Abbas, Mustafa Yavuz, and Bo~Cui.
\newblock Dry thermal development of negative electron beam resist polystyrene.
\newblock {\em Advances in Nano Research}, 1(2):105--109, June 2013.
\newblock Number: 2.

\bibitem{dey_effect_2013}
Ripon~Kumar Dey and Bo~Cui.
\newblock Effect of molecular weight distribution on e-beam exposure properties of polystyrene.
\newblock {\em Nanotechnology}, 24(24):245302, May 2013.

\bibitem{gupta_time-resolved_1982}
Mool~C. Gupta, Amitava Gupta, James Horwitz, and David Kliger.
\newblock Time-resolved fluorescence and emission depolarization studies on polystyrene: photochemical processes in polymeric systems. 9.
\newblock {\em Macromolecules}, 15(5):1372--1376, September 1982.

\bibitem{lee_photoluminescence_2006}
Hyeok~Moo Lee and Sung~Oh Cho.
\newblock Photoluminescence of the electron irradiated {Polystyrene}.
\newblock pages 1CD--ROM, Korea, Republic of, 2006. KNS.

\bibitem{lee_liquid_2007}
Hyeok~Moo Lee and Sung Oh~Cho Cho.
\newblock Liquid and {Solid}-{State} {NMR} study of the electron irradiated {Polystyrene}.
\newblock Korea, Republic of, 2007. KNS.

\bibitem{lee_fabrication_2008}
Hyeok~Moo Lee, Yong~Nam Kim, Bong~Hoon Kim, Sang~Ouk Kim, and Sung~Oh Cho.
\newblock Fabrication of {Luminescent} {Nanoarchitectures} by {Electron} {Irradiation} of {Polystyrene}.
\newblock {\em Advanced Materials}, 20(11):2094--2098, 2008.

\bibitem{kamura_fabrication_2018}
Yoshio Kamura and Kohei Imura.
\newblock Fabrication method of two-photon luminescent organic nano-architectures using electron-beam irradiation.
\newblock {\em Applied Physics Letters}, 112(24):243104, June 2018.

\bibitem{kamura_space-selective_2019}
Yoshio Kamura and Kohei Imura.
\newblock Space-{Selective} {Fabrication} of {Light}-{Emitting} {Carbon} {Dots} in {Polymer} {Films} {Using} {Electron}-{Beam}-{Induced} {Chemical} {Reactions}.
\newblock {\em Acs Omega}, 4(2):3380--3384, February 2019.

\bibitem{alexander_radiation_1997}
P.~Alexander, A.~Charlesby, and Francis~Arthur Freeth.
\newblock Radiation protection in copolymers of isobutylene and styrene.
\newblock {\em Proceedings of the Royal Society of London. Series A. Mathematical and Physical Sciences}, 230(1180):136--145, January 1997.

\bibitem{charlesby_swelling_1953}
A.~Charlesby.
\newblock Swelling properties of polystyrene crosslinked by high energy radiation.
\newblock {\em Journal of Polymer Science}, 11(6):521--529, 1953.

\bibitem{koike_radiation_1960}
Mitsuru Koike and Akibumi Danno.
\newblock Radiation {Effects} on {Dimethyl}-diphenyl {Siloxane} {Copolymer}. {I}. {Protective} {Effect} of {Phenyl} {Radical} on the {Cross}-linking.
\newblock {\em Journal of the Physical Society of Japan}, 15(8):1501--1508, August 1960.

\bibitem{burlant_-radiation_1962}
W.~Burlant, J.~Neerman, and V.~Serment.
\newblock γ-radiation of p-substituted polystyrenes.
\newblock {\em Journal of Polymer Science}, 58(166):491--500, 1962.

\bibitem{delides_protective_1980}
C.~G. Delides.
\newblock The protective effect of phenyl group on the crosslinking of irradiated dimethyldiphenylsiloxane.
\newblock {\em Radiation Physics and Chemistry (1977)}, 16(5):345--352, January 1980.

\bibitem{pankratova_effect_2000}
L.~N. Pankratova, L.~T. Bugaenko, and A.~A. Revina.
\newblock Effect of aromatic protectors on the radiolysis of polyorganosiloxanes.
\newblock {\em High Energy Chemistry}, 34(1):16--22, January 2000.

\bibitem{randall_13c_1983}
J.~C. Randall, F.~J. Zoepfl, and Joseph Silverman.
\newblock A {13C} {NMR} study of radiation-induced long-chain branching in polyethylene.
\newblock {\em Die Makromolekulare Chemie, Rapid Communications}, 4(3):149--157, 1983.

\bibitem{horii_carbon-13_1990}
Fumitaka Horii, Qingren Zhu, Ryozo Kitamaru, and Hitoshi Yamaoka.
\newblock Carbon-13 {NMR} study of radiation-induced crosslinking of linear polyethylene.
\newblock {\em Macromolecules}, 23(4):977--981, February 1990.

\bibitem{noauthor_radiation_1990}
{\em Radiation {Damage} to {Organic} {Materials} in {Nuclear} {Reactors} and {Radiation} {Environments} ({Proceedings} of a {Final} {Research} {Co}-ordination {Meeting}, {Takasaki}, {Japan}, 17-20 {July} 1989)}.
\newblock Number 551 in {TECDOC} {Series}. INTERNATIONAL ATOMIC ENERGY AGENCY, Vienna, 1990.

\bibitem{jahan1993effect}
MS~Jahan, DR~Ermer, and DW~Cooke.
\newblock Effect of x irradiation on optical properties of teflon-af.
\newblock {\em Radiation Physics and Chemistry}, 41(3):481--486, 1993.

\bibitem{forsythe1999radiation}
John~S Forsythe, David~JT Hill, Anestis~L Logothetis, and Andrew~K Whittaker.
\newblock The radiation chemistry of the copolymer of tetrafluoroethylene with 2, 2-bis (trifluoromethyl)-4, 5-difluoro-1, 3-dioxole.
\newblock {\em Polymer degradation and stability}, 63(1):95--101, 1999.

\bibitem{sultan_altering_2019}
Mansoor~A. Sultan, Sarah~K. Lami, Armin Ansary, Douglas~R. Strachan, Joseph~W. Brill, and J.~Todd Hastings.
\newblock Altering the radiation chemistry of electron-beam lithography with a reactive gas: a study of {Teflon} {AF} patterning under water vapor.
\newblock {\em Nanotechnology}, 30(30):305301, May 2019.

\bibitem{kumar_effect_2023}
Deepak Kumar, Krishnaroop Chaudhuri, Joseph~W. Brill, Jonathan~T. Pham, and J.~Todd Hastings.
\newblock Effect of water vapor pressure on positive and negative tone electron-beam patterning of poly(methyl methacrylate).
\newblock {\em Journal of Vacuum Science \& Technology B}, 41(1):012604, January 2023.

\bibitem{pivin_photoluminescence_2000}
J.~C Pivin, M~Sendova-Vassileva, P~Colombo, and A~Martucci.
\newblock Photoluminescence of composite ceramics derived from polysiloxanes and polycarbosilanes by ion irradiation.
\newblock {\em Materials Science and Engineering: B}, 69-70:574--577, January 2000.

\bibitem{huth2012focused}
Michael Huth, Fabrizio Porrati, Christian Schwalb, Marcel Winhold, Roland Sachser, Maja Dukic, Jonathan Adams, and Georg Fantner.
\newblock Focused electron beam induced deposition: A perspective.
\newblock {\em Beilstein journal of nanotechnology}, 3(1):597--619, 2012.

\bibitem{krysmann2012formation}
Marta~J Krysmann, Antonios Kelarakis, Panagiotis Dallas, and Emmanuel~P Giannelis.
\newblock Formation mechanism of carbogenic nanoparticles with dual photoluminescence emission.
\newblock {\em Journal of the American Chemical Society}, 134(2):747--750, 2012.

\bibitem{song2015investigation}
Yubin Song, Shoujun Zhu, Shitong Zhang, Yu~Fu, Li~Wang, Xiaohuan Zhao, and Bai Yang.
\newblock Investigation from chemical structure to photoluminescent mechanism: a type of carbon dots from the pyrolysis of citric acid and an amine.
\newblock {\em Journal of Materials Chemistry C}, 3(23):5976--5984, 2015.

\bibitem{zhu2013highly}
Shoujun Zhu, Qingnan Meng, Lei Wang, Junhu Zhang, Yubin Song, Han Jin, Kai Zhang, Hongchen Sun, Haiyu Wang, and Bai Yang.
\newblock Highly photoluminescent carbon dots for multicolor patterning, sensors, and bioimaging.
\newblock {\em Angewandte Chemie International Edition}, 52(14):3953--3957, 2013.

\bibitem{sun2013hair}
Dong Sun, Rui Ban, Peng-Hui Zhang, Ge-Hui Wu, Jian-Rong Zhang, and Jun-Jie Zhu.
\newblock Hair fiber as a precursor for synthesizing of sulfur-and nitrogen-co-doped carbon dots with tunable luminescence properties.
\newblock {\em Carbon}, 64:424--434, 2013.

\bibitem{zhang2016effect}
Yi~Zhang, Yaling Wang, Xiaoting Feng, Feng Zhang, Yongzhen Yang, and Xuguang Liu.
\newblock Effect of reaction temperature on structure and fluorescence properties of nitrogen-doped carbon dots.
\newblock {\em Applied Surface Science}, 387:1236--1246, 2016.

\bibitem{gedeon1999fast}
Ondrej Gedeon, Karel Jurek, and V{\'a}clav Hul{\i}́nsk{\`y}.
\newblock Fast migration of alkali ions in glass irradiated by electrons.
\newblock {\em Journal of non-crystalline solids}, 246(1-2):1--8, 1999.

\bibitem{gedeon2000microanalysis}
Ondrej Gedeon, V{\'a}clav Hul{\'\i}nsk{\`y}, and Karel Jurek.
\newblock Microanalysis of glass containing alkali ions.
\newblock {\em Microchimica Acta}, 132:505--510, 2000.

\bibitem{ghosh_photoluminescence_2014}
Siddharth Ghosh, Anna~M. Chizhik, Narain Karedla, Mariia~O. Dekaliuk, Ingo Gregor, Henning Schuhmann, Michael Seibt, Kai Bodensiek, Iwan A.~T. Schaap, Olaf Schulz, Alexander~P. Demchenko, Jörg Enderlein, and Alexey~I. Chizhik.
\newblock Photoluminescence of {Carbon} {Nanodots}: {Dipole} {Emission} {Centers} and {Electron}–{Phonon} {Coupling}.
\newblock {\em Nano Letters}, 14(10):5656--5661, October 2014.

\bibitem{dong_carbon-based_2013}
Yongqiang Dong, Hongchang Pang, Hong~Bin Yang, Chunxian Guo, Jingwei Shao, Yuwu Chi, Chang~Ming Li, and Ting Yu.
\newblock Carbon-{Based} {Dots} {Co}-doped with {Nitrogen} and {Sulfur} for {High} {Quantum} {Yield} and {Excitation}-{Independent} {Emission}.
\newblock {\em Angewandte Chemie}, 125(30):7954--7958, 2013.

\bibitem{nie_carbon_2014}
Hui Nie, Minjie Li, Quanshun Li, Shaojun Liang, Yingying Tan, Lan Sheng, Wei Shi, and Sean Xiao-An Zhang.
\newblock Carbon {Dots} with {Continuously} {Tunable} {Full}-{Color} {Emission} and {Their} {Application} in {Ratiometric} {pH} {Sensing}.
\newblock {\em Chemistry of Materials}, 26(10):3104--3112, May 2014.

\bibitem{wen_intrinsic_2013}
Xiaoming Wen, Pyng Yu, Yon-Rui Toh, Xiaotao Hao, and Jau Tang.
\newblock Intrinsic and {Extrinsic} {Fluorescence} in {Carbon} {Nanodots}: {Ultrafast} {Time}-{Resolved} {Fluorescence} and {Carrier} {Dynamics}.
\newblock {\em Advanced Optical Materials}, 1(2):173--178, 2013.

\bibitem{yu_temperature-dependent_2012}
Pyng Yu, Xiaoming Wen, Yon-Rui Toh, and Jau Tang.
\newblock Temperature-{Dependent} {Fluorescence} in {Carbon} {Dots}.
\newblock {\em The Journal of Physical Chemistry C}, 116(48):25552--25557, December 2012.

\bibitem{das_single-particle_2014}
Somes~K. Das, Yiyang Liu, Sinhea Yeom, Doo~Young Kim, and Christopher~I. Richards.
\newblock Single-{Particle} {Fluorescence} {Intensity} {Fluctuations} of {Carbon} {Nanodots}.
\newblock {\em Nano Letters}, 14(2):620--625, February 2014.

\bibitem{kim_white_2013}
Eunkyeom Kim, Jihoon Kyhm, Jung~Hyuk Kim, Gi~Yong Lee, Doo-Hyun Ko, Il~Ki Han, and Hyungduk Ko.
\newblock White light emission from polystyrene under pulsed ultra violet laser irradiation.
\newblock {\em Scientific Reports}, 3(1):3253, November 2013.

\bibitem{danilatos_foundations_1988}
G.~D. Danilatos.
\newblock Foundations of {Environmental} {Scanning} {Electron} {Microscopy}.
\newblock In Peter~W. Hawkes, editor, {\em Advances in {Electronics} and {Electron} {Physics}}, volume~71, pages 109--250. Academic Press, January 1988.

\bibitem{toth_effects_2000}
M.~Toth and M.~R. Phillips.
\newblock The effects of space charge on contrast in images obtained using the environmental scanning electron microscope.
\newblock {\em Scanning}, 22(5):319--325, 2000.

\bibitem{cummings1989charging}
KD~Cummings and M~Kiersh.
\newblock Charging effects from electron beam lithography.
\newblock {\em Journal of Vacuum Science \& Technology B: Microelectronics Processing and Phenomena}, 7(6):1536--1539, 1989.

\bibitem{he_measurement_2003}
Jing He and David~C. Joy.
\newblock Measurement of total gas scattering cross-section.
\newblock {\em Scanning}, 25(6):285--290, 2003.

\bibitem{rattenberger_method_2009}
J.~Rattenberger, J.~Wagner, H.~Schröttner, S.~Mitsche, and A.~Zankel.
\newblock A method to measure the total scattering cross section and effective beam gas path length in a low-vacuum {SEM}.
\newblock {\em Scanning}, 31(3):107--113, 2009.

\bibitem{nagasawa_charge_2008}
K.~Nagasawa, R.~Watanabe, Y.~Tanaka, and T.~Takada.
\newblock Charge accumulation in election beam irradiated various polymers.
\newblock In {\em 2008 {International} {Symposium} on {Electrical} {Insulating} {Materials} ({ISEIM} 2008)}, pages 147--150, Yokkaichi, September 2008. IEEE.

\bibitem{nagasawa_charge_2010}
Kenichiro Nagasawa, Masato Honjoh, Hiroaki Miyake, Rikio Watanabe, Yasuhiro Tanaka, and Tatsuo Takada.
\newblock Charge {Accumulation} in {Various} {Electron}-{Beam}-{Irradiated} {Polymers}.
\newblock {\em IEEJ Transactions on Electrical and Electronic Engineering}, 5(4):410--415, 2010.

\bibitem{ito}
Ito indium tin oxide coated glass /pet substrates, https://www.msesupplies.com/collections/ito-indium-tin-oxide-glass-substrates.

\bibitem{a2024_materials}
Materials database, https://tpsx.arc.nasa.gov/materialsdatabase, 2024.

\bibitem{ashida2009thermal}
Toru Ashida, Amica Miyamura, Nobuto Oka, Yasushi Sato, Takashi Yagi, Naoyuki Taketoshi, Tetsuya Baba, and Yuzo Shigesato.
\newblock Thermal transport properties of polycrystalline tin-doped indium oxide films.
\newblock {\em Journal of applied physics}, 105(7), 2009.

\bibitem{schram_ionization_1965}
B.L. Schram, F.J. De~Heer, M.J. Van Der~Wiel, and J.~Kistemaker.
\newblock Ionization cross sections for electrons (0.6–20 {keV}) in noble and diatomic gases.
\newblock {\em Physica}, 31(1):94--112, January 1965.

\bibitem{schutten_gross_1966}
J.~Schutten, F.~J. de~Heer, H.~R. Moustafa, A.~J.~H. Boerboom, and J.~Kistemaker.
\newblock Gross‐ and {Partial}‐{Ionization} {Cross} {Sections} for {Electrons} on {Water} {Vapor} in the {Energy} {Range} 0.1–20 {keV}.
\newblock {\em The Journal of Chemical Physics}, 44(10):3924--3928, May 1966.

\end{thebibliography}

\end{document}